\documentstyle[aps]{revtex}
\begin{document}
\draft \twocolumn[\hsize\textwidth\columnwidth\hsize\csname
@twocolumnfalse\endcsname

\title{Relational Quantum Mechanics}
\author{Carlo Rovelli}
\address{Department of Physics and Astronomy, 
University of Pittsburgh, Pittsburgh, Pa 15260, USA}
\date{\today}

\maketitle

\begin{abstract}
I suggest that the common unease with taking quantum mechanics as 
a fundamental description of nature (the measurement problem) could 
derive from the use of an incorrect notion, as the unease with the 
Lorentz transformations before Einstein derived from the notion of 
observer-independent time.  I suggest that this incorrect notion that 
generates the unease with quantum mechanics is the notion of 
observer-independent state of a system, or observer-independent 
values of physical quantities.  I reformulate the problem of the 
interpretation of quantum mechanics as the problem of deriving the 
formalism from a set of simple physical postulates.  I consider a 
reformulation of quantum mechanics in terms of information theory.  
All systems are assumed to be equivalent, there is no 
observer-observed distinction, and the theory describes only the 
information that systems have about each other; nevertheless, the 
theory is complete.
\end{abstract}
\vskip1cm

]

\section{A reformulation of the problem of the Interpretation of 
Quantum Mechanics}

In this paper, I discuss a novel view of quantum mechanics.  This 
point of view is not antagonistic to current ones, as the 
Copenhagen [Heisenberg 1927, Bohr 1935], consistent histories 
[Griffiths 1984, Griffiths 1996, Omnes 1988, Gell-Mann and Hartle 
1990], many-worlds [Everett 1957, Wheeler 1957, 
DeWitt 1970], quantum event [Huges 1989], many minds [Albert and 
Lower 1988, 1989, Lockwood 1986, Donald 1990] or modal [Shimony 
1969, van Fraassen 1991, Fleming 1992] interpretations, but rather 
combines and complements aspects of them.  This paper is based on 
a 
critique of a notion generally assumed uncritically.  As such, it 
bears a vague resemblance with Einstein's discussion of special  
relativity, which is based on the critique of the notion of absolute 
simultaneity.  The notion rejected here is the notion of absolute, or 
observer-independent, state of a system; equivalently, the notion of 
observer-independent values of physical quantities.  The thesis of 
the 
present work is that by abandoning such a notion (in favor of the 
weaker notion of state --and values of physical quantities-- {\it 
relative\/} to something), quantum mechanics makes much more 
sense.  
This conclusion derives from the observation that the experimental 
evidence at the basis of quantum mechanics forces us to accept that 
distinct observers give different descriptions of the same events.  
From this, I shall argue that the notion of observer-independent 
state 
of a system is inadequate to describe the physical world beyond the 
$\hbar \rightarrow 0$ limit, in the same way in which the notion of 
observer-independent time is inadequate to describe the physical 
world 
beyond the $c \rightarrow\infty$ limit.  I then consider the 
possibility of replacing the notion of absolute state with a notion 
that refers to the relation between physical systems.

The motivation for the present work is the commonplace observation 
that in spite of the 70 years-lapse from the discovery of quantum 
mechanics, and in spite of the variety of approaches developed with 
the aim of clarifying its content and improving the original 
formulation, quantum mechanics still maintains a remarkable level 
of 
obscurity. It is even accused of being unreasonable and unacceptable, 
even inconsistent, by world-class physicists (For example [Newman 
1993]).  My point of view in this regard is that quantum mechanics 
synthesizes most of what we have learned so far about the physical 
world:  The issue is thus not to replace or fix it, but rather to 
understand {\it what\/} precisely it says about the world; or, 
equivalently: what precisely we have learned from expe\-ri\-mental 
micro-physics.

It {\it is\/} difficult to overcome the sense of unease that quantum 
mechanics communicates.  The troubling aspect of the theory 
assumes 
different faces within different interpretations, and a 
complete description of the problem can only be based on a survey of 
the solutions proposed.  Here, I do not attempt such a survey; 
for a classical review see [d'Espagnat 1971], a more modern survey 
is 
in the first chapters of [Albert 1992], or, in compact form, see 
[Butterfield 1995].  The unease is expressed, for instance, in the 
objections the supporters of each interpretation raise against other 
interpretations.  Some of these objections are perhaps 
naive or ill-posed, but the fact that 
that no interpretation has so far succeeded in convincing the 
majority of 
the physicists, indicates, I believe, that the problem of the 
interpretation of quantum mechanics has not been fully disentangled 
yet.  This unease, and the variety of interpretations of quantum 
mechanics that it has generated is sometimes denoted as the 
measurement problem.  In this paper, I address this problem  
and consider a way out.

The paper is based on two ideas:  
\begin{itemize}
\item  That the unease may derive from the use of a concept which 
is 
inappropriate to describe the physical world at the quantum level. I 
shall argue that this concept is the concept of observer-independent 
state of a system, or, equivalently, the concept of 
observer-independent values of physical quantities.  
\item  That quantum mechanics will cease to look puzzling only 
when 
we will be able to {\it derive\/} the formalism of the theory from a 
set of simple physical assertions (``postulates",  ``principles") about 
the world.  Therefore, we should not try to  {\it append \/} a 
reasonable interpretation to the quantum mechanics  {\it 
formalism\/},  but rather to  {\it derive\/} the formalism from a set 
of experimentally motivated postulates. 
\end{itemize}
The reasons for exploring such a strategy are illuminated by an 
obvious historical precedent: special relativity. I shall make use of 
this analogy for explanatory purposes, in spite of the evident limits 
of the simile.

Special relativity is a well understood physical theory, 
appropriately credited to Einstein's 1905 celebrated paper. The 
formal content of special relativity, however, is coded into 
the Lorentz transformations, written by Lorentz, not by 
Einstein, and before 1905.  So, what was Einstein's 
contribution?  It was to understand the physical meaning of the 
Lorentz transformations. (And more, but this is what is of interest 
here). We could say --admittedly in a provocative manner-- that 
Einstein's contribution to special relativity has been the 
interpretation of the theory, not its formalism: the formalism 
already existed. 

Lorentz transformations were discussed at the beginning of the 
century, and their  {\it interpretation\/}  was debated. In 
spite of the recognized fact that they represent an extension of the 
Galilean group compatible with Maxwell theory, the Lorentz 
transformation were perceived as ``unreasonable'' and ``unacceptable 
as a fundamental spacetime symmetry'', even 
``inconsistent'', before 1905; words that recall nowadays comments 
on 
quantum mechanics.  The physical interpretation proposed by 
Lorentz himself (and defended by Lorentz long after 1905) was a 
physical contraction of moving bodies, caused by a complex and 
unknown 
electromagnetic interaction between the atoms of the bodies and the 
ether. It was quite an unattractive interpretation, remarkably 
similar to certain interpretations of the wave function collapse 
presently investigated!   Einstein's 1905 paper suddenly clarified 
the matter by pointing out the reason for the unease in taking  
Lorentz transformations ``seriously": the implicit use of a concept 
(observer-independent time) inappropriate to describe reality when 
velocities are high. Equivalently: a common deep assumption about 
reality (simultaneity is observer-independent) which is physically 
untenable.  The unease with the Lorentz transformations derived 
from 
a conceptual scheme in which an  incorrect notion  --absolute 
simultaneity-- was assumed, yielding any sort of paradoxical 
consequences. Once this notion was removed the physical 
interpretation 
of the Lorentz transformations stood clear, and special 
relativity is now considered rather uncontroversial.

Here I consider the hypothesis that all ``paradoxical" situations 
associated with quantum mechanics --as the famous and unfortunate 
half-dead Schr\"{o}dinger cat [Schr\"{o}dinger 1935]-- may derive 
from 
some 
analogous incorrect notion that we use in thinking about quantum 
mechanics. (Not in \textit{using} quantum mechanics, since we seem 
to have 
learned to use it in a remarkably effective way.)   The aim of this 
paper is to hunt for this incorrect notion, with the hope 
that by exposing it clearly to public contempt, we could free 
ourselves from the present unease with our best present theory of 
motion, and fully understand what does the theory assert about the 
world.  

Furthermore, Einstein was so persuasive with his interpretation of 
the 
Lorentz equations because he did not append an interpretation to 
them: 
rather, he re-derivedÊ them, starting from two postulates with 
terse 
physical content --equivalence of inertial observers and 
universality 
of the speed of light-- taken as facts of experience.  It was this 
re-derivation that unraveled the physical content of the Lorentz 
transformations and provided them a convincing interpretation.  I 
would like to suggest here that in order to clarify the physical 
meaning of quantum mechanics, a similar result should be searched: 
Finding a small number of simple statements about nature --which 
may perhaps seem contradictory, as the two postulates of special 
relativity do-- with clear physical content, from which the 
formalism of quantum mechanics could be derived.  In other words, I 
have a methodological suggestion for the problem of the 
interpretation of 
quantum mechanics: Finding the set of physical facts from which the 
quantum mechanics's formalism can be derived.  To my knowledge, 
such a derivation has not been achieved yet.  In this paper, I do not 
achieve such a result in a satisfactory manner, but I discuss a 
possible reconstruction scheme.

The program outlined is thus to do for the formalism of quantum 
mechanics what Einstein did for the Lorentz transformations:  i. Find 
a set of simple assertions about the world, with clear physical 
meaning, that we know are experimentally true (postulates); ii. 
Analyze these postulates, and show that from their conjunction it 
follows that certain common assumptions about the world are 
incorrect; iii. Derive the full formalism of quantum mechanics from 
these postulates. I expect that if this program could be completed, 
we would at long last begin to agree that we have understood 
quantum mechanics. 

In section 2, I analyze the measurement process as described by two 
distinct observers. This analysis leads to the main idea: the 
observer dependence of state and physical quantities, and to 
recognize a few key concepts in terms of which, I would like to 
suggest, the quantum mechanical description of reality ``makes 
sense".  Prominent among these is the concept of information 
[Shannon 
1949, Wheeler 1988, 1989, 1992].  In section 3, I switch from an 
inductive to a (very mildly) deductive mode, and put forward a set 
of notions, and a set of simple physical statements, from which the 
formalism of quantum mechanics can be reconstructed. I denote 
these 
statements as postulates, at the risk of misunderstanding: I do not 
claim any mathematical nor philosophical rigor, nor completeness 
 --supplementary assumptions are made along the way.  I 
am not interested here in a formalization of the subject, but only in 
better grasping its physics.  Ideas and techniques for the 
reconstruction are borrowed from quantum logic research, but 
motivations and spirit are different.  Finally, in section 4, I 
discuss the picture of the physical world that has emerged, and 
attempt an evaluation.  In particular, I compare the approach I have 
developed with some currently popular interpretations of quantum 
mechanics, and argue that the differences between these disappear, 
if 
the results presented here are taken into account.  

In order to prevent the reader from channeling his/her thoughts in 
the 
wrong direction, let me anticipate a few terminological remarks.  By 
using the word ``observer" I do not make any reference to conscious, 
animate, or computing, or in any other manner special, system.  I use 
the word ``observer" in the sense in which it is conventionally used 
in Galilean relativity when we say that an object has a velocity 
``with respect to a certain observer".  The observer can be any 
physical object having a definite state of motion.  For instance, I 
say that my hand moves at a velocity $v$ with respect to the lamp 
on 
my table.  Velocity is a relational notion (in Galilean as well as in 
special relativistic physics), and thus it is always (explicitly or 
implicitly) referred to something; it is traditional to denote this 
something as the observer, but it is important in the following 
discussion to keep in mind that the observer can be a table lamp.  
Also, I use information theory in its information-theory meaning 
(Shannon): information is a measure of the number of states in which 
a 
system can be --or in which several systems whose states are 
physically constrained (correlated) can be.  Thus, a pen on my table 
has information because it points in this or that direction.  We do 
not need a human being, a cat, or a computer, to make use of this 
notion of information.

\section{Quantum mechanics is a theory about information}

In this section, a preliminary analysis of the process of 
measurement 
is presented, and the main ideas are introduced. 
Throughout this section, standard quantum mechanics and standard 
interpretation --by which I mean for instance: formalism and 
interpretation in [Dirac 1930] or [Messiah 1958]-- are assumed.

\subsection{The third person problem}

Consider an observer $O$ (Observer) that makes a measurement on a 
system $S$ (System). For the moment we may think of $O$ as a 
classical 
macroscopic measuring apparatus, including or not including a human 
being.   Assume that the 
quantity being measured, say $q$, takes two values, 1 and 2; and let 
the states of the system $S$ be described by vectors (rays) in a two 
(complex) dimensional Hilbert space $H_S$.  Let the two eigenstates 
of the operator corresponding to the measurement of $q$ be 
$|1\rangle$ and $|2\rangle$.  As it is well known: if $S$ is in a 
generic normalized state  $|\psi\rangle =\alpha |1\rangle + \beta 
|2\rangle$, where $\alpha$ and $\beta$ are complex numbers and 
$|\alpha|^2 + |\beta|^2 = 1$, then $O$ can measure either one of the 
two values 1 and 2 -- with respective probabilities $|\alpha|^2$ and 
$|\beta|^2$.  

Assume that in a {\it given specific measurement\/}  the 
outcome of the measurement is 1.  From now on, we concentrate on 
describing this specific experiment, which we denote as $\cal E$.  
The system $S$ is affected by the measurement, and at a time 
$t=t_2$  
after the measurement, the state of the system is $|1\rangle$. In the 
physical sequence of events $\cal E$, the states of the system at 
$t_1$  and $t_2$ are thus
\begin{eqnarray}
	t_1        &	\longrightarrow	&	t_2   \nonumber \\
\alpha |1\rangle + \beta |2\rangle    &	\longrightarrow
	&|1\rangle	
\label{uno}
\end{eqnarray}

Let us now consider this same sequence of events $\cal E$, as 
described 
by a second  observer, which we refer to as $P$.  
I shall refer to $O$ as ``he" and to $P$ as ``she".  $P$ 
describes the interacting system formed by $S$ and $O$.  
Again, assume $P$ uses conventional quantum mechanics. Also, 
assume that 
$P$ does not perform any measurement on the $S-O$ system during 
the 
$t_1-t_2$ interval, but that she knows the initial states of both $S$ 
and $O$, and is thus able to give a quantum mechanical description 
of 
the set of events $\cal E$.  She describes the system $S$ by means 
of 
the Hilbert space $H_S$ considered above, and $O$ by means of a 
Hilbert 
space $H_O$.   The $S-O$ system is then described by the tensor 
product 
$H_{SO} = H_S\otimes H_O$.  As it has become conventional, let us 
denote 
the vector in $H_O$ that describes the state of the observer $O$ at 
$t=t_1$ (prior to the measurement) as $|init\rangle$.   The physical 
process during which $O$ measures the quantity $q$ of the system 
$S$ 
implies a physical interaction between $O$ and $S$. In the process 
of 
this interaction, the state of $O$ changes.  If the initial state of $S$ 
is $|1\rangle$ (resp $|2\rangle$) (and the initial state of $O$ is 
$|init\rangle$), then $|init\rangle$ evolves to a state that we 
denote as $|O1\rangle$ (resp $|O2\rangle$).  Think of $|O1\rangle$ 
(resp $|O2\rangle$) as a state in which ``the position of the hand of 
a measuring apparatus points towards the mark `1' (resp `2')". It is 
not difficult to construct model Hamiltonians that produce 
evolutions 
of this kind, and that can be taken as models for the physical 
interactions that produce a measurement.  Let us consider the actual 
case of the experiment $\cal E$, in which the initial state of $S$ is 
$|\psi\rangle = \alpha |1\rangle+\beta |2\rangle$. The initial full 
state of the $S-O$ system is then $|\psi\rangle\otimes |init\rangle 
= 
(\alpha |1\rangle+\beta |2\rangle)\otimes|init\rangle$. As well 
known, 
the linearity of quantum mechanics implies
\begin{eqnarray}
	t_1        &	\longrightarrow	&	t_2   \nonumber \\
	(\alpha |1\rangle + \beta |2\rangle)\otimes |init\rangle     
&	\longrightarrow	&
    \alpha |1\rangle\otimes |O1\rangle + \beta  |2\rangle\otimes 
|O2\rangle      	   \label{due}
\end{eqnarray}
Thus, at $t=t_2$Ê the system $S-O$ is in the state $(\alpha 
|1\rangle\otimes 
|O1\rangle+\beta|2\rangle\otimes|O2\rangle)$. This is the 
conventional 
description of a measurement as a physical process [von Neumann 
1932].  

I have described an actual physical process $\cal E$ taking place in a 
real laboratory.  Standard quantum mechanics requires us to 
distinguish system from observer, but it allows us freedom in 
drawing 
the line that distinguishes the two.  In the above analysis this 
freedom has been exploited in order to describe the same sequence 
of 
physical events in terms of two different descriptions.  In the first 
description, equation (1), the line that distinguishes system from 
observer is set between $S$ and O. In the second, equation (2), 
between $S-O$ and $P$.  Recall that we have assumed that $P$ is not 
making a measurement on the $S-O$ system; there is no physical 
interaction between $S-O$ and $P$ during the $t_1-t_2$ interval.  
$P$ 
may make measurements at a later time $t_{3}$: if she measures the 
value of $q$ on $S$ and the position of the hand on $O$, she finds 
that the two agree, because the first measurement collapses the 
state 
into one of the two factors of (2), leaving the second measurement 
fully determined to be the consistent value.  Thus, we have two 
descriptions of the physical sequence of events $\cal E$: The 
description (1) given by the observer $O$ and the description (2) 
given by the observer $P$.  These are two distinct {\it correct\/} 
descriptions of the same sequence of events $\cal E$.  At time 
$t_2$, 
in the $O$ description, the system $S$ is in the state $|1\rangle$ 
and 
the quantity $q$ has value 1.  According to the $P$ description, $S$ 
is not in the state $|1\rangle$ and the hand of the measuring 
apparatus does not indicate `1'.

Thus, I come to the observation on which the rest of the paper relies.
\begin{quote}
 {\bf Main observation:} In quantum mechanics different observers 
may give different accounts of the same sequence of events.
\end{quote}

For a very similar conclusion, see [Zurek 1982] and [Kochen 1979].  In 
the rest of the work, I explore the consequences of taking this 
observation fully into account.  Since this observation is crucial, I 
now pause to discuss and refute various objections to the 
main observation.  The reader who finds the above observation 
plausible may skip this rather long list of objections and jump to 
section II.C.

\subsection{Objections to the main observation}

{\it Objection 1.  Whether the account (1) or the account (2) is 
correct depends on which kind of system $O$ happens to be. There 
are 
systems that induce the collapse of the wave function. For instance, 
if $O$ is macroscopic (1) is correct, if $O$ is microscopic (2) is 
correct. }

The derivation of (2) does not rely on any assumption on the 
systems, but only from the basics of quantum mechanics (linearity).  
Therefore a particular $O$ system yielding (1) instead of (2) via 
Schr\"odinger evolution must behave in a way that contradicts the 
formalism of quantum mechanics as we know it.  This implies that 
$O$ 
cannot be described as a genuine quantum system.  Namely that there 
are special systems that do not obey conventional quantum 
mechanics, 
but are intrinsically classical in that they produce collapse of the 
wave functions --or the actualization of quantities' values.  This 
idea underlies a variety of old and recent attempts to unravel the 
quantum puzzle.  The special systems being for instance gravity 
[Penrose 1989], or minds [Albert and Loewer 1988], or macroscopic 
systems [Bohr 1949].  If we accept this idea, we have to separate 
reality into two kinds of systems: quantum mechanical systems on 
the 
one hand, and special systems on the other.  Bohr declares explicitly 
that we must renounce giving a quantum description of the classical 
world [Bohr 1949].  This is echoed in texts as [Landau and Lifschit 
1977].  Wigner pushes this view to the extreme consequences and 
distinguishes material systems (observed) from consciousness 
(observer) [Wigner 1961].  Here, on the contrary, I wish to assume

\begin{quote}
{\bf Hypothesis 1}: All systems are equivalent: Nothing distinguishes 
a 
priori macroscopic systems from quantum systems.  If the 
observer $O$ can give a quantum description of the system $S$, then 
it is 
also legitimate for an observer $P$ to give a quantum description of 
the system formed by the observer $O$.  
\end{quote}

Of course, I have no proof of hypothesis 1, only plausibility 
arguments.  I am suspicious toward attempts to introduce special 
non-quantum and not-yet-understood new physics, in order to 
alleviate 
the strangeness of quantum mechanics: they look to me very much 
like 
Lorentz' attempt to postulate a mysterious interaction that 
Lorentz-contracts physical bodies ``for real" --something that we 
now 
see was very much off of the point, in the light of Einstein's 
clarity.  Virtually all those views modify quantum mechanical 
predictions, in spite of statements of the contrary: if at $t_2$, the 
state is as in (1), then $P$ can never detect interference terms 
between the two branches in (2), contrary to quantum theory 
predictions.  These discrepancies are likely to be minute, as shown 
by 
the beautiful discovery of the physical mechanism of decoherence 
[Zurek 1981, Joos and Zeh 1985], which ``saves the phenomena".  But 
they are nevertheless different from zero, and thus observable (more 
on this later).  I am inclined to trust that a sophisticated 
experiment able to detect those minute discrepancies will fully 
vindicate quantum mechanics against distortions due to postulated 
intrinsic classicality of specific systems.  In any case, the question 
is experimentally decidable; and we shall see.  Second, I do not like 
the idea that the present over-successful theory of motion can only 
be 
understood in terms of its failures yet-to-be-detected.  Finally, I 
think it is reasonable to remain committed, up to compelling 
disproof, 
to the rule that all physical systems are equivalent with respect to 
mechanics: this rule has proven so successful, that I would not 
dismiss it as far as there is another way out.

\vskip.2cm 

{\it Objection 2. What the discussion indicates is that the quantum 
state is different in the two accounts, but the quantum state is a 
fictitious non-physical mental construction; the physical content of 
the theory is given by the outcomes of the measurements.  }  

Indeed, one can take the view that outcomes of measurements are 
the 
physical content of the theory, and the quantum state is a 
secondary theoretical construction. This is the way I read 
[Heisenberg 1927] and  [van Fraassen 1991].  According to this view, 
anything in between two measurement outcomes is like the 
``non-existing" trajectory of the electron, to use Heisenberg's vivid 
expression, of which there is nothing to say.  I am very sympathetic 
with this view, which plays an important role in section III.  This 
view, however, does not circumvent the main observation for the 
following reason. The account (2) states that there is nothing to be 
said about the value of the quantity $q$ of $S$ at time $t_2$:  for 
$P$, 
at $t= t_2$  the quantity $q$ does not have a determined value.  On 
the 
other hand, for $O$, at $t= t_2$, $q$ has value 1. From which the 
main observation follows again. 

\vskip.2cm 

{\it Objection 3. As before (only outcomes of measurements are 
physical), but the truth of the matter is that $P$ is right and $O$ is 
wrong. } 

This is indefensible. Since all physical experiments of which we 
know 
can be seen as instances of the $S-O$ measurement, this would 
imply 
that not a single outcome of measurement has ever being obtained 
yet. 
If so, how could have we learned quantum theory?

\vskip.2cm

{\it Objection 4. As before (only outcomes of measurements are 
physical), but the truth of the matter is that $O$ is right and 
P is wrong.} 

If $P$ is wrong, quantum mechanics cannot be applied to the $S-O$ 
system 
(because her account is a straightforward implementation of 
textbook 
quantum mechanics). Thus this objection predicts discrepancies, so 
far never observed, with quantum mechanical predictions, which 
include observable interference effects between the two terms of 
(2).

\vskip.2cm 

{\it Objection 5. As before, but under the assumption that $O$ is 
macroscopic. Then interference terms become extremely 
small because of decoherence effects.  If they are small enough, they 
are unobservable, and thus $q=1$ becomes an absolute property of 
$S$, 
which is true and absolutely determined, albeit unknown to $P$, who 
could measure it anytime, and would not see interference effects.} 

Strictly speaking this is wrong, because decoherence depends 
on which observation $P$ will make. Therefore, the property 
$q=1$ of $S$ would become an absolute property at time $t_2$, or 
not, 
according to which subsequent properties of $S$ the observer $P$ 
considers. This is the reason for which the idea of exploiting 
physical decoherence for the basic interpretation  of 
quantum mechanics problem has evolved 
into the consistent histories interpretations, where 
probabilities are (consistently) assigned to histories, and not to 
single outcomes of measurements within a history. (See, however, 
the 
discussion on the no-histories slogan in [Butterfield 1995].)

\vskip.2cm 

{\it Objection 6.  There is no collapse. The description (1) is not 
correct, because ``the wave function never really collapses". The 
account  (2)  is the correct one. There are no values assigned to 
classical properties of system; there are only quantum states.}   

If so, then the observer $P$ cannot measure the value of the property 
$q$ either, since (by assumption) there are no values assigned to 
classical properties, but only quantum states; thus the quantity $q$ 
doesn't ever have a value.  But we do describe the world in terms of 
``properties" that the systems have and values assumed by various 
quantities, not in terms of states in Hilbert space.  In a description 
of the world purely in terms of quantum states, the systems never 
have 
definite properties and I do not see how to match such a description 
with {\it any\/} observation.  For a detailed elaboration of this 
point, which is too often neglected, but I think is very strong, see 
[Albert 1992].

\vskip.2cm 

{\it Objection 7.  There is no collapse. The description (1) is not 
correct, because ``the wave function never really collapses". The 
account  (2)  is the correct one. The values assigned to classical 
properties are different from branch to branch.} 

This is a form of Everett's view [Everett 1957], which entails the 
idea that when we measure the electron's spin being up, the electron 
spin is also and simultaneously down ``in some other branch" --or 
``world", hence the many world denomination of this view.   The 
property of the electron of having spin up is not absolutely 
true, but only true relative to ``this" branch. We have 
a new ``parameter" for expressing contingency: ``which branch" is a 
new 
``dimension" of indexicality, in addition to the familiar ones 
``which time" and ``which place".  Thus, the state of affairs of the 
example is that, at $t_2$, $q$ has value 1 in one branch and has 
value 2 
in the other; the two branches being theoretically described by the 
two terms in (2).  This is a fascinating idea that has recently been 
implemented in a variety of diverse incarnations. Traditionally, the 
idea has been discussed in the context of a notion of apparatus, 
namely a distinguished set of subsystems of the universe, and a 
distinguished quantity of such an apparatus --the preferred basis. 
Such a (collection of) preferred apparatus and preferred basis are 
needed in order to define branching, and thus in order to have 
assignment of values [Butterfield 1995]; the view has recently 
branched into the many mind interpretations, where the 
distinguished subsystems are related to various aspect of the human 
brain. (See [Butterfield 1995] for a recent discussion). These 
versions of Everett's idea violate hypothesis 1, and thus I am not 
concerned with them.  Alternatively, there are versions of 
Everett's idea that reject the specification of preferred apparatus 
and preferred basis, and in which the branching itself is indexed by 
an arbitrarily chosen system playing the role of apparatus and an 
arbitrarily chosen basis. To my knowledge, the only elaborated 
versions of this view which avoids the difficulties mentioned in 
objection 5 have evolved into the histories formalisms considered 
below. 

\vskip.2cm 

{\it Objection 8.  What is absolute and observer independent is the 
probability of a sequence  $A_1, ... A_n$ of property ascriptions 
(such 
that the interference terms mentioned above are extremely small - 
decoherence);  this probability is independent from the existence of 
any observer measuring these properties. } 

This is certainly correct.  In fact, this observation is at the root 
of the consistent histories (CH) interpretations of quantum 
mechanics 
[Griffiths 1984, 1996, Omnes 1988, Gell-Mann and Hartle 1990].  
However, in my understanding, CH confirms the observation above 
that 
different observers give different accounts of the same sequence of 
events, for the following reason.  The beauty of the histories 
interpretations is the fact that the probability of a sequence of 
events in a consistent family of sequences does not depends on the 
observer, precisely as it doesn't in classical mechanics.  One can be 
content with this powerful result of the theory and stop here.  
However, probabilities depends on the choice of the consistent {\it 
family\/} of histories, which is chosen (to avoid misunderstanding: 
{\it whether or not\/} a physical occurrence can be assigned a 
probability depends on the family chosen).  One (who?)  makes a 
choice 
in picking up a family of alternative histories in terms of which he 
chooses to describe the system.  Griffiths has introduced the vivid 
expression ``framework'' to indicate a consistent family of histories 
[Griffiths 1996].  There exist funny cases in which one framework is: 
``Is the value of the physical quantity $Q$ equal to 1 at time $t$, or 
not?"; and a second framework is ``Is the value of the physical 
quantity $Q$ equal to 2 at time $t$, or not?", and in each framework 
the answer is yes with probability 1!  [Kent 1995, Kent and Dowker 
1995, Griffiths 1996.]  Therefore the description of what has 
happened 
at time $t$, that we can give on the basis of a fixed set of data, is: 
At $t$, $Q$ was equal to 1, if we ask whether $Q$ was 1 or not.  Or: 
At $t$, $Q$ was equal to 2, if we ask whether $Q$ was 2 or not.  
There 
is no contradiction here (in Copenhagen terms, the same 
mathematics 
would indicate that the outcome depends on which apparatus is 
present), but it is difficult to deny that a large majority of 
physicists still want to understand more about this strangeness of 
quantum mechanics.  In the Copenhagen view, the choice that 
corresponds  
to the choice between frameworks is determined by which classical 
apparatus is present.  Namely, the framework is determined by the 
interaction of the quantum system with a classical object.  In CH, 
one 
claims that property ascription does not need a classical 
interaction; 
the price to pay is that (probabilistic) predictions, rather than 
being uniquely determined, are framework dependent.  In the example 
of 
the previous section, the observers $O$ and $P$ may choose two 
distinct 
frameworks, and the corresponding two descriptions are both valid: 
each one in its own framework.  However, observer $O$ does not have 
the choice of using the framework that the observer $P$ uses, 
because 
he has ``seen'' $q=1$.  After having seen that $q=1$, $O$ has no 
option anymore of allowing a framework in which $q$ is not 1.  The 
fact that $q=1$ has become one of his ``dataÕÕ; and data {\it 
determine\/} which frameworks are consistent.  Therefore, the two 
observers $O$ and $P$ have different sets of frameworks at their 
disposal for describing the same events, because they have different 
data (for the same set of events).  The framework in which (2) 
makes 
sense is available to $P$, but not to $O$, because $O$ has data that 
include that fact that $Q=1$ at $t_2$.  What is {\it data\/} for $O$ 
is not data for $P$, who considers the full $S-O$ system: $P$ is still 
allowed to choose a framework which does not include the value 1 of 
$q$ at $t_2$.  Once the data are specified, all predictions are well 
defined in CH, but the characterization of what may count as data, 
and 
therefore which frameworks are available, is different for the two 
observers of the above example.  Once more, we have that two 
different 
observers give different descriptions of the same set of events: 
what 
is data for $O$ is only a possible choice of framework for $P$.  I 
will return on this delicate point in the last section.

\vskip.2cm 

In conclusion, it seems to me that whatever view of quantum theory 
(consistent with hypothesis 1) one holds, the main observation is 
inescapable.  I may thus proceed to the main point of this work.   

\subsection{Main discussion}

If different observers give different accounts of the same sequence 
of events, then each quantum mechanical description has to be 
understood as relative to a particular observer.  Thus, a quantum 
mechanical description of a certain system (state and/or values of 
physical quantities) cannot be taken as an ``absolute" (observer 
independent) description of reality, but rather as a formalization, 
or codification, of properties of a system {\it relative\/}  to a 
given observer. Quantum mechanics can therefore be viewed as a 
theory 
about the states of systems and values of physical quantities 
relative to other systems.  

A quantum description of the state of a system $S$ exists only if 
some 
system $O$ (considered as an observer) is actually ``describing" $S$, 
or, 
more precisely, has interacted with $S$. The quantum state of a 
system 
is always a state of that system with respect to a certain other 
system. More precisely: when we say that a physical quantity takes 
the value $v$, we should always (explicitly or implicitly) qualify 
this statement as: the physical quantity takes the value $v$  with 
respect to the so and so observer.  Thus, in the example considered 
in section 2.1, $q$ has value 1 {\it with respect to \/} $O$, but not 
with respect to $P$.
 
Therefore, I suggest that in quantum mechanics ``state" as well as 
``value of a variable" --or ``outcome of a measurement--" are 
relational 
notions in the same sense in which velocity is relational in 
classical mechanics.  We say ``the object $S$ has velocity $v$'' 
meaning ``with respect to a reference object $O$". Similarly, I 
maintain that 
``the system is in such a quantum state" or ``$q=1$" are always to be 
understood ``with respect to the reference $O$."  In quantum 
mechanics 
{\it all\/} physical variables are relational, as is velocity.    

If quantum mechanics describes relative information only, one could 
consider the possibility that there is a deeper underlying theory  
that describes what happens ``in reality".  This is the thesis of the 
incompleteness of quantum mechanics (first suggested in [Born 
1926]!). Examples of hypothetical underlying theories are hidden 
variables theories [Bohm 1951, Belifante 1973]. Alternatively, the 
``wave-function-collapse-producing" systems can be ``special" 
because 
of some non-yet-understood physics, which becomes relevant due to 
large number of degrees of freedom [Ghirardi Rimini and Weber 
1986, 
Bell 1987], complexity [Hughes 1989], quantum gravity [Penrose 
1989] 
or other.  

As is well known, there are no indications on {\it physical\/}  
grounds that quantum mechanics is incomplete. Indeed, the {\it 
practice\/} of quantum mechanics supports the view that quantum 
mechanics represents the best we can say about the world at the 
present state of experimentation, and suggests that the structure of 
the world grasped by quantum mechanics is deeper, and not 
shallower, 
than the scheme of description of the world of classical mechanics. 
On the other hand, one could consider motivations on {\it 
metaphysical\/}  grounds, in support of the incompleteness of 
quantum 
mechanics. One could argue: ``Since reality has to be real and 
universal, and the same for everybody, then a theory in which the 
description of reality is observer-dependent is certainly an 
incomplete theory".  If such a theory were complete, our concept of 
reality would be disturbed.  

But the way I reformulated the problem of the interpretation of 
quantum mechanics in section I.~should make us suspicious and 
attentive 
precisely to such kinds of arguments, I believe.  Indeed, what we are 
looking for is precisely some ``wrong general assumption" that we 
suspect to have, and that could be at the origin of the unease with 
quantum mechanics.  Thus, I discard the thesis of the 
incompleteness of quantum mechanics and assume

\begin{quote}
{\bf Hypothesis 2} (Completeness): Quantum mechanics provides a 
complete and self-consistent scheme of description of the physical 
world, appropriate to our present level of experimental 
observations. 
\end{quote}
The conjunction of this hypothesis 2 with the main observation of 
section II.A and the discussion above leads to the following idea:
\begin{quote}
{\it Quantum mechanics is a theory about the physical description of 
physical systems relative to other systems, and this is a complete 
description of the world.} 
   \end{quote}
The thesis of this paper is that this conclusion is not 
self-contradictory. If this conclusion is valid, then the incorrect 
notion at the source of our unease with quantum theory has been 
uncovered: it is the notion of true, universal, observer-independent 
description of the state of the world.  If the notion of 
observer-independent description of the world is unphysical, a  
complete description of the world is exhausted by the relevant 
information that systems have about each other.  Namely, there is 
neither an absolute state of the system, nor absolute properties that 
the system has at a certain time.  Physics is fully relational, not 
just as far as the notions of rest and motion are considered, but 
with respect to all physical quantities.  Accounts (1) and (2) of the 
sequence of events $\cal E$ are both correct, even if distinct: any 
time we talk about a state or property of a system, we have to 
refer these notions to a specific observing, or reference system. 
Thus, I propose the idea that quantum mechanics indicates that the 
notion of a universal description of the state of the world, shared 
by all observers, is a concept which is physically untenable, on 
experimental ground.\footnote{To counter objections based on 
instinct 
alone, it is perhaps worthwhile recalling the great resistance that 
the idea of fully relational notions of ``rest" and ``motion" 
encountered at the beginning of the scientific revolution. I think that 
quantum mechanics 
(and general relativity) could well be in the course of triggering  
a --not yet developed-- revision of world views as far reaching as 
the 
seventeenth century's one (on this, see [Rovelli 1995]).}

Thus, the hypothesis on which I base this paper is that accounts (1) 
and (2)  are both fully correct. They refer to different observers.  
I propose to reinterpret 
every contingent statement about nature (``the 
electron has spin up", ``the atom is in the so and so excited state", 
the ``spring is compressed", ``the chair is here and not there") as 
elliptic expressions for relational assertions (``the electron has 
spin up {\it with respect to the Stern Gerlac apparatus\/}" ... ``the 
chair is here and not there {\it with respect to my eyes\/}", and so 
on).    A general physical theory is 
a theory about the state that physical systems have, relative to each 
other.  I explore and elaborate this possibility in this paper. 

\subsection{Relation between descriptions}

The multiplication of points of view induced by the 
relational notion of state and physical quantities' values considered 
above raises the problem of the relation between distinct 
descriptions of 
the same events.  What is the relation between the value of a 
variable 
$q$ relative to an observer $O$, and the value of the same variable 
relative to a different observer?  This problem is subtle.  Consider 
the example of section II.A. We expect some relation between the 
description of the world illustrated in (1) and in (2).

First of all, one may ask what is the ``actual'', ``absolute'' 
relation between the description of the world relative to $O$ and the 
one relative to $P$.  This is a question debated in the context of 
``perspectival'' interpretations of quantum mechanics.  I think 
that the question is ill-posed.  The absolute state of affairs of the 
world is a meaningless notion; asking about the absolute relation 
between two descriptions is 
precisely asking about such an absolute state of affairs of the 
world.  
Therefore there is no meaning in the ``absolute'' relation between the 
views of different observers.  In particular, there is no way of 
deducing the view of one from the view of the other.

Does this mean that there is no relation whatsoever between views 
of 
different observers?  Certainly not; it means that the relation itself 
must be understood quantum mechanically rather than classically.  
Namely the issue of the relation between views must be addressed 
within the view of one of the two observers (or of a third one).  In 
other words, we may investigate the view of the world of $O$, as 
seen 
by $P$.  Still in other words: the fact that a certain quantity $q$ 
has a value with respect to $O$ is a physical fact; as a physical 
fact, its being true, or not true, must be understood as relative to 
an observer, say $P$.  Thus, the relation between $O$'s and $P$'s 
views is not absolute either, but it can be described in the 
framework of, say, $P$'s view.

There is an important physical reason behind this fact: It {\it is\/} 
possible to compare different views, but the process of comparison 
is always a physical interaction, and all physical interactions are 
quantum mechanical in nature.  I think that this simple fact is 
forgotten in most discussions on quantum mechanics, yielding 
serious 
conceptual errors.  Suppose a physical quantity $q$ has value with 
respect to you, as well as with respect to me.  Can we compare 
these 
values?  Yes we can, by communicating among us.  But 
communication is 
a physical interaction and therefore is quantum mechanical.  In 
particular, it is intrinsically probabilistic.  Therefore you can 
inquire about the value of $q$ with respect to me, but this is (in 
principle) a quantum measurement as well.

Next, one must distinguish between two different questions: (i)  
Does 
$P$ ``know" that $S$ ``knows" the value of $q$?  (ii)  Does $P$ know 
what is the value of $q$ relative to $O$?  (I know {\it that\/} you 
know the amount of your salary, but I do not know {\it what\/} you 
know about the amount of your salary).

(i) Can $P$ ``know" that $O$ has made a measurement on $S$ at time 
$t_2$?  The answer is yes.  $P$ has a full account of the events 
$\cal 
E$.  Description (2) expresses the fact that $O$ has measured $S$.  
The key observation is that in the state at $t_2$ in (2), the 
variables $q$ (with eigenstates $|1\rangle$ and $|2\rangle$) and the 
pointer variable (with eigenstates $|O1\rangle$ and $|O2\rangle$) 
are 
correlated.  From this fact, $P$ understands that the 
pointer variable in $O$ has information about $q$.  In fact, the state 
of $S-O$ is the quantum superposition of two states: in the first, 
($|1\rangle\otimes |O1\rangle$), $S$ is in the $|1\rangle $ state and 
the hand of the observer is correctly on the `1' mark.  In the second, 
($|2\rangle\otimes |O2\rangle$), $S$ is in the $|2\rangle$ state and 
the hand of the observer is, correctly again, on the `2' mark.  In 
both cases, the hand of $O$ is on the mark that correctly represents 
the state of the system.  More formally, there is an operator $M$ on
the Hilbert space of the $S-O$ system whose physical interpretation
is ``Is the pointer correctly correlated to $q$?''
If  $P$ measures $M$, then the outcome of this 
measurement would be yes with certainty, when the state of the 
$S-O$ system is as in (2).  The operator $M$ is given by
\begin{eqnarray}
	M \ (|1\rangle \otimes |O1\rangle)& = &|1\rangle\otimes 
|O1\rangle 
\nonumber \\
 	M \ (|1\rangle\otimes |O2\rangle)& = & 0   
 \nonumber \\
	M \ (|2\rangle\otimes |O2\rangle)& = & |2\rangle\otimes 
|O2\rangle 	
\nonumber \\	
M  \ ( |2\rangle\otimes |O1\rangle ) & = &   0 	           
\end{eqnarray}
where the eigenvalue 1 means ``yes, the hand of $O$ indicates the 
correct state of S" and the eigenvalue 0 means ``no, the hand of $O$ 
does not indicate the correct state of S".  At time $t_2$, the $S-O$ 
system is in an eigenstate of $M$ with eigenvalue 1; therefore $P$ 
can 
predict with certainty that $O$ ``knows'' the value of $q$.  Thus, it 
is meaningful to say that, according to the $P$ description of the 
events $\cal E$, $O$ ``knows" the quantity $q$ of $S$, or that he 
``has measured" the quantity $q$ of $S$, and the pointer variable 
embodies the information.

A side remark is important.  In general, the state of the $S-O$ 
system 
will not be an eigenvalue of $M$.  In particular, the physical 
interaction between $S$ and $O$ which establishes the correlation 
will 
take time.  Therefore the correlation between the $q$ variable of 
$S$ 
and the pointer variable of $O$ will be established gradually.  Does 
this mean that, in $P$ views, the measurement is made ``gradually'', 
namely that, according to $P$, $q$ will have value with respect to 
$O$ 
only partially?  This is a much debated question: ``Half the way 
through the measurement, has a measurement being done?''.  By 
realizing that $P$'s knowledge about $O$ is also quantum 
mechanical, 
we find --I believe-- the solution of the puzzle: If the state of the 
$S-O$ system is not an eigenstate of $M$, then, following standard 
quantum mechanics rule, this means that any eventual attempt of 
$P$ to 
verify whether or not a measurement has happened will have 
outcome 
``yes'' or ``no'' with a certain respective probability.  In other 
words: there is no half-a-measurement; there is probability one-half 
that the measurement has been made!  We never see quantum 
superpositions of physical values, we only see physical values, but 
we 
can predict which one we are going to see only probabilistically.  
Similarly, I can say only probabilistically whether or not a physical 
quantity has taken value for you; but I should not say that you 
``half-see'' a physical quantity!  Thus, by representing the fact that 
(for $P$) ``the pointer variable of $O$ has information about the $q$ 
variable in $S$" by means of the operator $M$ resolves the well-
known 
and formidable problem of defining the ``precise moment" in which 
the 
measurement is performed, or the precise ``amount of correlation" 
needed for a measurement to be established --see for instance 
[Bacciagaluppi and Hemmo 1995].  Such questions are not classical 
questions, but quantum mechanical questions, because whether or 
not 
$O$ has measured $S$ is not an absolute property of the $S-O$ state, 
but a quantum property of the quantum $S-O$ system, that can be 
investigated by $P$, and whose yes/no answers are, in general, 
determined only probabilistically.  In other words: {\it imperfect 
correlation does not imply no measurement performed, but only a 
smaller than 1 probability that the measurement has been 
completed.  }

A second remark in this regard is that, due to the well-known 
bi-orthogonal decomposition theorem, there are always correlated 
variables in any coupled system (in a pure state).  Therefore there is 
always ``some'' operator $M$ for which the $S-O$ system is an 
eigenstate.  Much emphasis has been given to this fact in the 
literature.  I do not think this fact is very relevant.  Imagine we 
have a quantum particle in a box, with a finite probability to tunnel 
out of it (say this models a nuclear  decay).  At some initial time we 
describe the state with a wave function concentrated in the box.  At 
some later time a Geiger counter detects the particle outside the 
box, 
and we describe the particle as a position eigenstate at the Geiger 
counter position.  During the time in between, we can describe the 
state of the particle by giving the wave form of its Schr\"odinger 
wave $\psi(x)$ as it leaks out of the box.  Now, in principle, we know 
that there is an operator $A$ in the Hilbert space of the particle 
such that $\psi(x)$ is an eigenstate of $A$.  Therefore we know that 
``some'' quantity is uniquely defined at any moment.  But what is the 
interest of such observation?  Very little, I would say.  $A$ will 
correspond to some totally uninteresting and practically non 
measurable quantity.  Similarly, given an arbitrary state of the 
coupled $S-O$ system, there will always be a basis in each of the 
two 
Hilbert spaces which gives the bi-orthogonal decomposition, and 
therefore which defines an $M$ for which the coupled system is an 
eigenstate.  But this is of null practical nor theoretical 
significance.  We are interested in {\it certain\/} self-adjoint 
operators only, representing observables that we know how to 
measure; 
for this same reason, we are only interested in correlations between 
{\it certain\/} quantities: the ones we know how to measure.

The second question $P$ may ask is: (ii.)  What is the outcome of the 
measurement performed by $O$?  It is important not to confuse the 
statement ``$P$ knows {\it that\/} $O$ knows the value of $q$" with 
the statement ``$P$ knows {\it what\/} $O$ knows about $q$".  In 
general, the observer $P$ does not know ``what is the value of the 
observable $q$ that $O$ has measured" (unless $\alpha$ or $\beta$ 
in 
(2) vanish).  An observer with sufficient initial information may 
predict which variable the other observer has measured, but not the 
outcome of the measurement.  Communication of measurements 
results is 
however possible (and fairly common!).  $P$ can measure the 
outcome of 
the measurement performed by $O$.  She can, indeed, measure 
whether 
$O$ is in $|O1\rangle$, or in $|O2\rangle$.  

Notice that there is a consistency condition to be fulfilled, which is 
the following: if $P$ knows that $O$ has measured $q$, then she 
measures $q$, and then she measures what $O$ has obtained in 
measuring 
$q$ (namely she measures the pointer variable), then consistency 
requires that the results obtained by $P$ on the $q$ variable and on  
the pointer variable be correlated.  Indeed, they are!  as was first 
noticed 
by von Neumann, and as is clear from (2).  Thus, there is a satisfied 
consistency requirement in the notion of relative description 
discussed.  This can be expressed in terms of standard quantum 
mechanical language: From the point of view of the $P$ description:
\begin{quote} 
{\it The fact that the pointer variable in $O$ has information about 
$S$ (has measured $q$) is expressed by the existence of a 
correlation 
between the $q$ variable of $S$ and the pointer variable of $O$.  The 
existence of this correlation is a measurable property of the $S-O$ 
state.}
\end{quote}

\subsection{Information}

It is time to introduce the main concept in terms of which I propose 
to interpret quantum mechanics: information.  

What is the precise nature of the relation between the variable $q$ 
and the system $O$ expressed in the statement ``$q=1$ relative to 
$O$"?  Does this relation have a comprehensible physical meaning?  
Can 
we analyze it in physical terms?  The answer has emerged in the 
previous subsection.  Let me recapitulate the main idea: The 
statement 
``$q$ has a value relative to $O$" refers to the contingent state of 
the $S-O$ system.  But the contingent state of the $S-O$ system has 
no 
observer-independent meaning.  We can make statements about the 
state 
of the $S-O$ system only provided that we interpret these 
statements 
as relative to a third physical system $P$.  Therefore, it should be 
possible to understand what is the physical meaning of ``$q$ has a 
value relative to $O$'' by considering the description that $P$ gives 
(or could give) of the $S-O$ system.  This description is not in terms 
of classical physics, but in quantum mechanical terms; it is the one 
given in detail above.  The result is that ``$q$ has value with 
respect to $S$ means that there is a correlation between the 
variable 
$q$ and the pointer variable in $O$, namely that $P$ is able to 
predict that subsequent measurements she will make on $q$ and on 
the
pointer variable will produce correlated outcomes. 

Correlation is ``information'' in the sense of information theory 
[Shannon 
1949]. If the state of the $S-O$ system is in an eigenstate of $M$ 
with
eigenvalue 1, then the four possible configurations that the $q$ 
variable
and the pointer variable can take are reduced to two. Therefore (by
definition) the pointer variable has information about $q$. Let me 
then
take a lexical move.  I will from now on express the 
fact that $q$ has a certain value with respect to $O$ by saying: $O$ 
has the ``information'' that $q=1$.  

The notion of information I employ here should not be confused with 
other notions of information used in other contexts.  I use here a 
notion of information that does not require distinction between 
human 
and non-human observers, systems that understand meaning or don't, 
very-complicated or simple systems, and so on.  As it is well known, 
the problem of defining such a notion was brilliantly solved by 
Shannon: in the technical sense of information-theory, the amount of 
information is the number of the elements of a set of alternatives 
out 
of which a configuration is chosen.  Information expresses the fact 
that a system is in a certain configuration, which is correlated to 
the configuration of another system (information source).  The 
relation between this notion of information and more elaborate 
notions 
of information is given by the fact that the information-theoretical 
information is a minimal condition for more elaborate notions.  In a 
physical theory it is sufficient to deal with this basic 
information-theoretical notion of information.  This is very weak; it 
does not require us to consider information storage, 
thermodynamics, 
complex systems, meaning, or anything of the sort.  In particular: 
(i.)  information can be lost dynamically (correlated state may 
become 
uncorrelated); (ii.)  we do not distinguish between correlation 
obtained on purpose and accidental correlation; Most important: (iii.)  
any physical system may contain information about another physical 
system.  For instance if we have two spin-1/2 particles that have 
the 
same value of the spin in the same direction, we say that one has 
information about the other one.  Thus observer system in this paper 
is any possible physical system (with more than one state).  If there 
is any hope of understanding how a system may behave as observer 
without renouncing the postulate that all systems are equivalent, 
then 
the same kind of processes --``collapse''-- that happens between an 
electron and a CERN machine, may also happen between an electron 
and 
another electron.  Observers are not ``physically special systems" in 
any sense.  The relevance of information theory for understanding 
quantum physics has been advocated by John Wheeler [Wheeler 1988, 
1989, 1992].

Thus, the physical nature of the relation between $S$ and $O$ 
expressed in the fact that $q$ has a value relative to $O$ is captured 
by the fact that $O$ has information (in the sense of information 
theory) about $q$.  By ``$q$ has a value relative to $O$", we mean 
``relative to $P$, there is a certain correlation in the $S$ and $O$ 
states", or, equivalently, ``O has information about $q$".

Notice that this is, in a sense, only a partial answer to the question 
formulated at the beginning of this section.  First, it is a quantum 
mechanical answer, because $P$'s information about the $S-O$ 
system is 
probabilistic.  Second, it is an answer that only shifts the problem 
by one step, because the information possessed by $O$ is explained 
in 
terms of the information possessed by $P$.  Thus, the notion of 
information I use has a double valence.  On the one hand, I want to 
weaken all physical statements that we make: not ``the spin is up", 
but ``we have information that the spin is up" --which leaves the 
possibility open to the fact that somebody other has different 
information.  Thus, {\it information\/} indicates the usual 
ascription 
of values to quantities that founds physics, but emphasizes their 
relational aspect.  On the other hand, this ascription can be 
described within the theory itself, as information-theoretical {\it 
information\/}, namely correlation.  But such a description, in turn, 
is quantum mechanical and observer dependent, because a universal 
observer-independent description of the state of affairs of the 
world 
does not exist.  Finally, there is a key irreducible distinction 
between $P$'s knowledge that $O$ has information about $q$ and 
$O$'s 
knowledge of $q$.  Physics is the theory of the relative information 
that systems have about each other.  This information exhausts 
everything we can say about the world.

At this point, the main ideas and concepts have been formulated.  In 
the next section, I consider a certain number of postulates 
expressed 
in terms of these concepts, and derive quantum mechanics from 
these 
postulates.

\section{On the reconstruction of Quantum Mechanics}

\subsection{ Basic concepts}

Physics is concerned with relations between physical systems.  In 
particular, it is concerned with the description that physical 
systems 
give of other physical systems.  Following hypothesis 1, I reject any 
fundamental distinctions as: system/observer, quantum/ classical 
system, physical system/consciousness.  Assume that the world can 
be 
decomposed (possibly in a variety of ways) in a collection of 
systems, 
each of which can be equivalently considered as an observing system 
or 
as an observed system.  A system (observing system) may have 
information about another system (observed system).  Information is 
exchanged via physical interactions.  The actual process through 
which 
information is collected and perhaps stored is not of particular 
interest here, but can be physically described in any specific 
instance.

Information is a discrete quantity: there is a minimum amount of 
information exchangeable (a single bit, or the information that 
distinguishes between just two alternatives.)  I will denote a 
process 
of acquisition of information (a measurement) as a ``question" that a 
system (observing system) asks another system (observed system).  
Since information is discrete, any process of acquisition of 
information can be decomposed into acquisitions of elementary bits 
of 
information.  I refer to an elementary question that collects a single 
bit of information as a ``yes/no question", and I denote these 
questions as $Q_1, Q_2, \ldots $.

Any system $S$, viewed as an observed system, is characterized by 
the 
family of yes/no questions that can be asked to it.  These correspond 
to the physical variables of classical mechanics and to the 
observables of conventional quantum mechanics.  I denote the set of 
these questions as $W(S) = \{Q_i, i \in I\}$, where the index $i$ 
belongs to an index set $I$ characteristic of $S$.  The general 
kinematical features of $S$ are representable as relations between 
the 
questions $Q_i$ in $W(S)$, that is, structures over $W(S)$.  For 
instance, meaningful questions that can be asked to an electron are 
whether the particle is in a certain region of space, whether its spin 
along a certain direction is positive, and so on.

The result of a sequence of questions $( Q_1, Q_2 , Q_3, \ldots )$ to 
$S$, from an observer system $O$, can be represented by a string
\begin{equation}
			(e_1, e_2, e_3, \ldots  )
\end{equation}
where each $e_i$ is either 0 or 1 (no or yes) and represents the 
response of the system to the question $Q_i$.  Thus the information 
that $O$ has about $S$ can be represented as a binary string.  It is a 
basic fact about nature that knowledge of a portion $(e_1, \ldots , 
e_n)$ of this string provides indications about the subsequent 
outcomes $(e_{n+1}, e_{n+2}, \ldots )$.  It is in this sense that a 
string (4) contains the information that $O$ has about $S$.

Trivially repeating the same question (experiment) and obtaining 
always the same outcome does not increase the information on $S$.  
The 
{\it relevant\/} information (from now on, simply information) that 
$O$ has about $S$ is defined as the non-trivial content of the 
(potentially infinite) string (4), that is the part of (4) relevant 
for predicting future answers of possible future questions.  The 
relevant information is the subset of the string (4), obtained 
discarding the $e_i$'s that do not affect the outcomes of future 
questions.

The relation between the notions introduced and traditional notions 
used in quantum mechanics is transparent: A question is a version of 
a 
measurement.  The idea that quantum measurements can be reduced 
to 
yes/no measurements is old.  A yes/no measurement is 
represented by a projection operator onto a linear subset of the 
Hilbert space, or by the linear subset of the Hilbert space itself.  
Here this idea is not derived from the quantum mechanical 
formalism, 
but is justified in information-theoretical terms.  The notions of 
observing system and observed system reflect the traditional 
notions 
of observer and system (but any system can play both roles here).  
$W(S)$ corresponds to the set of the observables.  Recall that in 
algebraic approaches a system is characterized by the (algebraic) 
structure of the family of its observables.

A notion does not appear here: the state of the system.  
The absence of this notion is the prime feature of the interpretation 
considered here.  In place of the notion of state, which refers solely 
to the system, the notion of the information that a system has about 
another system has been introduced.  I view this notion very 
concretely: a piece of paper on which outcomes of measurements are 
written, hands of measuring apparatus, memory of scientists, or a 
two-value variable which is up or down after an interaction.

For simplicity, in the following I focus on systems that in 
conventional quantum mechanics are described by a finite 
dimensional 
Hilbert space.  This choice simplifies the mathematical treatment of 
the theory, avoiding continuum spectrum and other infinitary issues.

\subsection{The two main postulates}

\begin{quote}
{\bf Postulate 1} (Limited information).  There is a maximum amount 
of 
relevant information that can be extracted from a system.  
\end{quote} 
The physical meaning of postulate 1 is that it is possible to exhaust, 
or give a complete description of the system.  In other words, any 
future prediction that can be inferred about the system out of an 
infinite string (4), can also be inferred from a finite subset 
\begin{equation} 
s = [e_1, \ldots , e_N] 
\end{equation} 
of (4), where $N$ is a number that characterizes the system $S$.  
The 
finite string (5) represents the maximal knowledge that $O$ has 
about 
$S$.\footnote{The string (5) is essentially the state.  The novelty 
here is not the fact that the state is defined as the response of the 
system to a set of yes/no experiments: this is the traditional 
reading 
of the state as a preparation procedure.  The novelty is that this 
notion of state is relative to the observer that has asked the 
questions.} One may say that any system $S$ has a maximal 
``information capacity" $N$, where $N$, an amount of information, is 
expressed in bits.  This means that $N$ bits of information exhaust 
everything we can say about $S$.  Thus, each system is 
characterized 
by a number $N$.  In terms of traditional notions, we can view $N$ as 
the smallest integer such that $N \geq \log_2 k$, where $k$ is the 
dimension of the Hilbert space of the system $S$.  Recall that the 
outcomes of the measurement of a complete set of commuting 
observables, characterizes the state, and in a system described by a 
$k = 2^N$ dimensional Hilbert space such measurements distinguish 
one 
outcome out of $2^N$ alternative (the number of orthogonal basis 
vectors): this means that one gains information $N$ on the system.  
Postulate 1 is confirmed by our experience about the world (within 
the 
assumption above, that we restrict to finite dimensional Hilbert 
space 
systems.  Generalization to infinite systems should not be difficult.)

Notice that postulate 1 already adds the Planck's constant to 
classical physics.  Consider a classical system described by a 
variable $q$ that takes bounded but continuous values; for instance, 
the position of a particle.  Classically, the amount of information we 
can gather about it is infinite: we can locate its state in the 
system's phase space with arbitrary precision.  Quantum 
mechanically, 
this infinite localization is impossible because of postulate 1.  
Thus, maximum available information can localize the state only 
within 
a finite region of the phase space.  Since the dimensions of the 
classical phase space of any system are $(L^{2}T^{-1}M)^n$, there 
must 
be a universal constant with dimension $L^{2}T^{-1}M$, that 
determines 
the minimal localizability of objects in phase space.  This constant 
is of course Planck's constant.  Thus we can view Planck's constant 
just as the transformation coefficient between physical units 
(position $\times$ momentum) and information theoretical units 
(bits).

What happens if, after having asked the $N$ questions such that the 
maximal information about $S$ has been gathered, the system $O$ 
asks a 
further question $Q_{N+1}$?
\begin{quote}
{\bf Postulate 2} (Unlimited information).  It is always possible to 
acquire new information about a system.
 \end{quote}
If, after having gathered the maximal information about $S$, the 
system $O$ asks a further question $Q$, to the observed system $S$, 
there are two extreme possibilities: either the question $Q$ is fully 
determined by previous questions, or not.  In the first case, no new 
information is gained.  However, the second postulate asserts that 
there is always a way to acquire new information.  This postulate 
implies therefore that the sequence of responses we obtain from 
observing a system cannot be fully deterministic.

The motivation for the second postulate is fully experimental.  We 
know that all quantum systems (and all systems are quantum 
systems) 
have the property that even if we know their quantum state 
$|\psi\rangle$ exactly, we can still ``learn" something new about 
them 
by performing a measurement of a quantity $O$ such that 
$|\psi\rangle$ 
is not an eigenstate of $O$.  This is an {\it experimental\/} result 
about the world, coded in quantum mechanics.  Postulate 2 expresses 
this result.

Since the amount of information that $O$ can have about $S$ is 
limited 
by postulate 1, when new information is acquired, part of the old 
relevant-information becomes irrelevant.  In particular, if a new 
question $Q$ (not determined by the previous information gathered), 
is 
asked, then $O$ looses (at least) one bit of the previous information.  
So that, after asking the question $Q$, new information is available, 
but the total amount of relevant information about the system does 
not 
exceed $N$ bits.

Rather surprisingly, those two postulates are (almost) sufficient to 
reconstruct the full formalism of quantum mechanics.  Namely, one 
may 
assert that the physical content of the general formalism of 
quantum 
mechanics is (almost) nothing but a sequence of consequences of 
two 
physical facts expressed in postulates 1 and 2.  This is illustrated 
in the next section.

\subsection{Reconstruction of the formalism, and the third 
postulate}

In this section, I discuss the possibility of deriving the formalism 
of quantum mechanics from the physical assertions contained in the 
postulates 1 and 2.  This section is technical, and the uninterested 
reader may skip it and jump to section III.D. The technical machinery 
I employ has been developed (with different motivations) in quantum 
logic analyses.  See for example [Beltrametti and Cassinelli 1981].  
As I mentioned in the introduction, the reconstruction attempt is not 
fully successful.  I will be forced to introduce a third postulate 
(besides various relative minor assumptions).  I will speculate on 
the 
possibility of giving this postulate a simple physical meaning, but I 
do not have any clear result.  This difficulty reflects parallel 
difficulties in the quantum logic reconstruction attempts.

Let me begin by analyzing the consequences of the first postulate.  
The number of questions in $W(S)$ can be much larger than $N$.  
Some 
of these questions may not be independent.  In particular, one may 
find (experimentally) that they can be related by implication 
$(Q_1\Rightarrow Q_2)$, union $(Q_3=Q_1\vee Q_2)$ and 
intersection 
$(Q_3=Q_1 \wedge Q_2)$.  One can define an always false $(Q_0)$ 
and an 
always true question $(Q_\infty)$, the negation of a question $(\neg 
Q)$, and a notion of orthogonality as follows: if $Q_1 \Rightarrow 
\neg Q_2$, then $Q_1$ and $Q_2$ are orthogonal (we indicate this as 
$Q_1 \bot Q_2$).  Equipped with these structures, and under the 
(non-trivial) additional assumption that $\vee$ and $\wedge$ are 
defined for every pair of questions, $W(S)$ is an orthomodular 
lattice 
[Beltrametti and Cassinelli 1981, Huges 1989].

If there is a maximal amount of information that can be extracted 
from 
the system, we may assume that one can select in $W(S)$ an 
ensemble of 
$N$ questions $Q_i$, which we denote as $c = \{Q_i, i=1, N\}$, that 
are independent from each other.  There is nothing canonical in this 
choice, so there may be many distinct families $c, b, d, ...$ of $N$ 
independent questions in $W(S)$.  If a system $O$ asks the $N$ 
questions in the family $c$ to a system $S$, then the answers 
obtained 
can be represented as a string that we denote as
\begin{equation}
			s_cÊ  =   [e_1, ...... , e_N]_c				      
\end{equation}
The string $s_c$ represents the information that $O$ has about $S$, 
as 
a result of the interaction that allowed it to ask the questions in 
$c$.  The string $s_c$ Êcan take $2^N = K$ values; we denote these 
values as $s_c^{(1)}, s_c^{(2)}, ...  , s_c^{(K)}$.  So that
\begin{eqnarray}
	s_c^{(1)} &=& [0, 0, \ldots , 0]_c 
	\nonumber \\  	
	s_c^{(2)} &=& [0, 0, \ldots , 1]_c  
	\nonumber \\  
	&\ldots&,  \nonumber \\  
	 s_c^{(K)} &=& [1, 1, \ldots , 1]_c  	
\end{eqnarray}
Since the $2^N$ possible outcomes $s_c^{(1)},Ês_c^{(2)}, ... ,s_c^{(K)}$ 
of 
the $N$ yes/no 
questions are (by construction) mutually exclusive, we can define 
$2^N$ new 
questions $Q_c^{(1)}... Q_c^{(K)}$ such that the yes answer to 
$Q_c^{(i)}$  
corresponds to the string of answers $s_c^{(i)}$:
\begin{eqnarray}
		 Q_c^{(1)} &=& \neg   Q_1  \wedge \neg   Q_2  \wedge ....  
\wedge 
\neg  Q_N \nonumber \\
Q_c^{(2)} &=& \neg Q_1 \wedge \neg Q_2 \wedge ....  \wedge Q_N 
\nonumber \\ &...& \nonumber \\
		 Q_c^{(k)} &=&   Q_1  \wedge   Q_2  \wedge ....  \wedge  
Q_N 
\end{eqnarray}
We refer to questions of this kind as ``complete questions".  By 
taking all possible unions of sets of complete questions $Q_c^{(i)}$ 
(of the same family $c$), we construct a Boolean algebra that has 
$Q_c^{(i)} $ as atoms.

Alternatively, the observer $O$ could use a different family of $N$ 
independent yes-no questions, in order to gather information about 
$S$.  Denote an alternative set as $b$.  Then, he will still have a 
maximal amount of relevant information about $S$ formed by an 
$N$-bit 
string $s_bÊ = [e_1, ......  , e_N]_b$.  Thus, $O$ can give different 
kinds of descriptions of $S$, by asking different questions.  
Correspondingly, denote as $s_b^{(1)}...  s_b^{(K)}$ the $2^N$ values 
that $s_b$ can take, and consider the corresponding complete 
questions 
$Q_b^{(1)}...  Q_b^{(K)}$ and the Boolean algebra they generate.  
Thus, it follows from the first postulate that the set of the 
questions $W(S)$ that can be asked to a system $S$ has a natural 
structure of an orthomodular lattice containing subsets that form 
Boolean algebras.  This is precisely the algebraic structure formed 
by 
the family of the linear subsets of a Hilbert space, which represent 
the yes/no measurements in ordinary quantum mechanics!  [Jauch 
1968, 
Finkelstein 1969, Piron, 1972, Beltrametti and Cassinelli 1981.]

The next question is the extent to which the information (6) about 
the 
set of questions c determines the outcome of an additional question 
$Q$.  There are two extreme possibilities: that $Q$ is fully 
determined by (6), or that it is fully independent, namely that the 
probability of getting a yes answer is 1/2.  In addition, there is a 
range of intermediate possibilities: The outcome of $Q$ may be 
determined probabilistically by $s_c$.  The second postulate states 
explicitly that there are questions that are non-determined.  Define, 
in general, as $p(Q, Q_c^{(i)})$ the probability that a yes answer to 
$Q$ will follow the string $s_c^{(i)}$.  Given two complete families 
of information $s_c$ and $s_b$, we can then consider the 
probabilities\footnote{I do not wish to enter here the debate on the 
meaning of probability in quantum mechanics.  I think that the shift 
of perspective I am suggesting is meaningful in the framework of an 
objective definition of probability, tied to the notion of repeated 
measurements, as well as in the context of subjective probability, 
or 
any variant of this, if one does not accept Jayne's criticisms of the 
last.}
\begin{equation}
 			p^{ij} =  p(Q_b^{(i)}, Q_c^{(j)}) 
\end{equation}
From the way it is defined, the $2^N \times 2^N$ matrix $p^{ij}$ 
cannot be 
fully 
arbitrary.  First, we must have
\begin{equation}
		0 \geq p^{ij} \geq    1 		
\end{equation}
Then, if the information $s_c^{(j)}$, is available about the system, 
one and only one of the outcomes $s_b^{(i)}$, may result.  Therefore
\begin{equation}
			\sum_i   p^{ij}  = 1						 
\end{equation}
We also assume that $p(Q_b^{(i)}, Q_c^{(j)}) = p(Q_c^{(j)},Q_b^{(i)})$ 
(this is a new assumption!  There is a relation with time reversal, 
but I leave it here as an unjustified assumption at this stage), from 
which we must have
\begin{equation}
			\sum_j   p^{ij}  = 1 					 
\end{equation}
The conditions (10-11-12) are strong constraints on the matrix 
$p^{ij}$.  They are satisfied if
\begin{equation}
			 p^{ij}   = | U^{ij} |^2						
\end{equation}
where $U$ is a unitary matrix, and $p^{ij}$ can always be written in 
this form for some unitary matrix $U$ (which, however, is not fully 
determined by $p^{ij}$).
 
Consider a question in the Boolean 
algebra generated by a family $s_c$, for instance   
\begin{equation}
			Q_c^{(jk)}  = Q_c^{(j)} \vee Q_c^{(k)}
\end{equation}
In order to take this question into account, 
we cannot consider probabilities of the form $p(Q_b^{(i)}, 
Q_c^{(jk)})$, because a 
yes answer to $Q_c^{(jk)}$ is less than the maximum amount of 
relevant information.  But we may consider probabilities of the 
form, say, 
\begin{equation}
		p^{i(jk)i} =  p(Q_b^{(i)}, Q_c^{(jk)} Q_b^{(i)})			
	\end{equation}
defined as the probability that a yes answer to $Q_b^{(i)}$ will 
follow a yes answer to $Q_b^{(i)}$ ($N$ bits of information) and a 
subsequent yes answer to $Q_c^{(jk)}$ ($N-1$ bits of information).  
As 
is well known, we have (experimentally!)  that
\begin{eqnarray}
p^{i(jk)i} &\neq &  
		p(Q_b^{(i)}, Q_c^{(j)})\ p(Q_c^{(j)},Q_b^{(i)}) 
		\nonumber \\
	&& 	+ p(Q_b{(i)},Q_c^{(k)})\ p(Q_c^{(k)},Q_b^{(i)}) \nonumber \\
& = &  (p^{ij})^2 +  (p^{ik})^2 
\end{eqnarray}
Accordingly, we can determine the missing phases of $U$ in (13) by 
means 
of 
the correct relation, which is
\begin{equation}
		p^{i(jk)i}  =  | U^{ij}U^{ji}  +  U^{ik}U^{ki} |^2			
\end{equation}
It would be extremely interesting to study the constraints that the 
probabilistic nature of all quantities $p$ implies, and to investigate 
to which extent the structure of quantum mechanics can be derived 
in 
full from these constraints.  One could conjecture that eqs.(13-17) 
could be derived solely by the properties of conditional probabilities 
--or find exactly the weakest formulation of the superposition 
principle directly in terms of probabilities: this would be a strong 
result.  Alternatively, it would be even more interesting to 
investigate the extent to which the noticed consistency between 
different observers' descriptions, which I believe characterizes 
quantum mechanics so marvelously, could be taken as the missing 
input 
for reconstructing the full formalism.  I have a suspicion this could 
work, but have no definite result.  Here, I content myself with the 
more modest step of introducing a third postulate.  For strictly 
related attempts to reconstruct the quantum mechanical formalism 
from 
the algebraic structure of the measurement outcomes, see [Mackey 
1963, 
Maczinski 1967, Finkelstein 1969, Jauch 1968, Piron 1972].
\begin{quote}
{\bf Postulate 3} (Superposition principle).  If $c$ and $b$ define 
two complete families of questions, then the unitary matrix 
$U_{cb}$ 
in
\begin{equation}
			p( Q_c^{(i)}, Q_b^{(j)} ) = |U_{cb}^{ij}|^{2} 	
\end{equation}
can be chosen in such a way that for every $c$, $b$ and $d$, we have 
$U_{cd} = U_{cb} U_{bd}$ and the effect of composite questions is 
given by eq.(17).
\end{quote}

It follows that we may consider any question as a vector in a 
complex 
Hilbert space, fix a basis $|Q_c^{(i)} \rangle $ in this space and 
represent any other question $|Q_b^{(j)} \rangle$ as a linear 
combination of these:
\begin{equation}
			|Q_b^{(j)} \rangle  =  \sum_i    U_{bc}^{ji}\ \ 
|Q_c^{(i)} 
\rangle  	
\end{equation}
The matrices $U_{bc}^{ij}$ are then a unitary change of basis from 
the 
$|Q_c^{(i)} \rangle$ to the $|Q_b^{(j)} \rangle$ basis.  Recall now 
the conventional quantum mechanical probability rule: if $|v^{(i)} 
\rangle$ are a set of basis vectors and $|w^{(j)} \rangle$ a second 
set of basis vectors related to the first ones by
\begin{equation}
	|w^{(j)} \rangle  =  \sum_i   U^{ji}\ |v^{(i)} \rangle  
\end{equation}
then the probability of measuring the state  $|w^{(j)} \rangle$  if the 
system is in the state   $|w^{(i)} \rangle$   is 
\begin{equation}
			p^{ij} = |  \langle  v^{(i)} | w^{(j)} \rangle  |^2
	\end{equation}
(20) and (21) yield $p^{ij} = |U^{ij}|^2$, which is equation (18).  
Therefore the conventional formalism of quantum mechanics as well 
as the 
standard probability rules follow completely from the three 
postulates.  
The set $W(S)$ has the structure of a set of linear subspaces in the 
Hilbert space.  For any yes/no question $Q_i$, let $L_i$ be the 
corresponding linear subset of $H$.  The relations $\{ \Rightarrow, 
\vee, \wedge, \neg, \bot \}$ between questions $Q_i$ correspond to 
the 
relations $\{$inclusion, orthogonal sum, intersection, 
orthogonal-complement, orthogonality$\}$ between the 
corresponding 
linear subspaces $L_i$. 

The inclusion of dynamics in the above scheme is straightforward.  
Two 
questions can be considered as distinct if defined by the same 
operations but performed at different times.  Thus, any question can 
be labeled by the time variable $t$, indicating the time at which it 
is asked: denote as $t \rightarrow Q(t)$ the one-parameter family of 
questions defined by the same procedure performed at different 
times.  
In this way we have naturally the Heisenberg picture.  As we have 
seen, the set $W(S)$ has the structure of a set of linear subspaces in 
the Hilbert space.  Assuming that time evolution is a symmetry in 
the 
theory, the set of all the questions at time $t_2$ must be 
isomorphic 
to the set of all the questions at time $t_1$.  Therefore the 
corresponding family of linear subspaces must have the same 
structure; 
therefore there should be a unitary transformation $U( t_2 - t_1 )$ 
such that
\begin{equation}
		Q( t_2 ) = U( t_2 - t_1 ) Q( t_1 ) U^{-1}( t_2 - t_1 )  
\end{equation}
By conventional arguments, these unitary matrices form an abelian 
group and $U( t_2 - t_1 ) = exp\{-\imath(t_2-t_1)H\}$, where $H$ is 
a 
self-adjoint operator on the Hilbert space, the Hamiltonian.  The 
Schr\"odinger equation follows immediately if we transform from 
the 
Heisenberg to the Schr\"odinger picture. 

\subsection{The observer observed }

We now have the full formal machinery of quantum mechanics, with 
an 
interpretative novelty: the absence of the notion of state the 
system.  
I now return to the issue of the relation between information of 
distinct observers.  How can a system $P$ have information about 
the 
fact that $O$ has information about $S$?  The information possessed 
by 
distinct observers cannot be compared directly.  This is the key 
point 
of the construction.  A statement {\it about\/} the information 
possessed by $O$ is a statement about the physical state of $O$; the 
observer $O$ is a regular physical system.  Since there is no 
absolute 
meaning to the state of a system, any statement regarding the state 
of 
$O$, including the information it possess, is to be referred to some 
other system observing $O$.  A second observer $P$ can have 
information about the fact that $O$ has information about $S$, but 
any 
acquisition of information implies a physical interaction.  $P$ can 
get new information about the information that $O$ has about $S$ 
only 
by physically interacting with the $S-O$ system.

At the cost of repeating myself, let me stress again that I believe 
that the common mistake in analyzing measurement issues in 
quantum 
mechanics is to forget that two observers can compare their 
information (their measurement outcomes) only by physically 
interacting with each other.  This means that there is no way to 
compare ``the information possessed by $O$'' with "the information 
possessed by $P$", without considering a {\it quantum\/} physical 
interaction, 
or a quantum measurement, between the two. 

The relation between the information possessed by distinct 
observers 
is thus given by the following: Viewed by $O$, information about $S$ 
is the primary concept in terms of which one describes the world; 
viewed by $P$, the information that $O$ has about $S$ information 
is 
just a property of some degrees of freedom in $O$ being correlated 
with some property of $S$.  This can be taken as an additional 
ingredient
to the structure defined by the three postulates; it ties the distinct 
observers to each other. 

Again, the direct question ``Do observers $O$ and $P$ have the same 
information on a system $S$?'' is meaningless, because it is a 
question about the absolute state of $O$ and $P$.  What is 
meaningful 
is to rephrase the question in terms of some observer.  For instance, 
we could ask it in terms of the information possessed by a further 
observer, or by $P$ herself.  Consider this last case.  At time $t_1$, 
$O$ gets information about $S$.  $P$ has information about the 
initial 
state, and therefore has the information that the measurement has 
been 
performed.  The meaning of this is that she knows that the states of 
the $S-O$ systems are correlated, or more precisely she knows that 
if 
at a later time $t_3$ she asks a question to $S$ concerning property 
$A$, and a question to $O$ concerning his knowledge about $A$ (or, 
equivalently, concerning the position of a pointer), she will get 
consistent results.

From the dynamical point of view, knowledge of the structure of the 
family of questions $W(S)$ implies the knowledge of the dynamics 
of 
$S$ (because $W(S)$ includes all Heisenberg observables at all 
times).  
In Hilbert space terms, this means knowing the Hamiltonian of the 
evolution of the observed system.  If $P$ knows the dynamics of the 
$S-O$ system, she knows the two Hamiltonians of $O$ and $S$ {\it 
and\/} the interaction Hamiltonian.  The interaction Hamiltonian 
cannot be vanishing because a measurement ($O$ measuring $S$) 
implies 
an interaction: this is the only way in which a correlation can be 
dynamically established.  From the point of view of $P$, the 
measurement is therefore a fully unitary evolution, determined by a 
peculiar interaction Hamiltonian between $O$ and $S$.  The 
interaction 
is a measurement if it brings the states (relative $P$) to a 
correlated configuration.  On the other hand, $O$ gives a dynamical 
description of $S$ alone.  Therefore he can only use the $S$ 
Hamiltonian.  Since between times $t_1$ and $t_2$ the evolution of 
$S$ 
is affected by its interaction with $O$, the description of the 
unitary evolution of $S$ given by $O$ breaks down.  The unitary 
evolution does not break down for mysterious physical quantum 
jumps, 
or due to unknown effects, but simply because $O$ is not giving a 
full 
dynamical description of the interaction.  $O$ cannot have a full 
description of the interaction of $S$ with himself ($O$), because his 
information is correlation, and there is no meaning in being 
correlated with oneself.

The reader may convince himself that even if we take into account 
several observers observing each other, there is no way in which 
contradiction may develop, provided that one does not violate the 
two 
rules:
\begin{itemize}
\item (i)  There is no meaning to the state of a system or the 
information that a system has, except within the information of a 
further observer. 
\item (ii) There is no way a system $P$ may get information about a 
system $O$ without physically interacting with it, and therefore 
without breaking down (at the time of the interaction) the unitary 
evolution description of $O$.  
\end{itemize}
For instance, there is no way two observers $P$ and $O$ can get 
information about a system $S$ independently from each other: one 
of 
two (say $O$) will have to obtain the information first.  In doing so, 
he will interact with $S$ at a certain time $t$.  This interaction 
implies that there is a non vanishing interaction Hamiltonian 
between 
$S$ and $O$.  If $P$ asks a question to $O$ at a later time $t'$, she 
will either have to consider the interacting correlated $S-O$ 
system, 
or to realize that the unitary evolution of the $O$ dynamics has 
broken down, due to the physical interaction she was not taking into 
account.

\section{Critique of the concept of state}

\subsection{``Any observation requires an observer'': summary of the 
ideas presented}

Let me summarize the path covered.  I started from the distinction 
between observer and observed-system.  I assumed (hypothesis 1) 
that 
all systems are equivalent, so that any observer can be described by 
the same physics as any other system.  In particular, I assumed that 
an observer that measures a system can be described by quantum 
mechanics.  I have analyzed a fixed physical sequence of events 
$\cal 
E$, from two different points of observations, the one of the 
observer 
and the one of a third system, external to the measurement.  I have 
concluded that two observers give different accounts of the same 
physical set of events (main observation).

Rather than backtracking in front of this observation, and giving up 
the commitment to the belief that all systems are equivalent, I have 
decided to take this experimental fact at its face value, and 
consider 
it as a starting point for understanding the world.  {\it If different 
observers give different descriptions of the state of the same 
system, 
this means that the notion of state is observer dependent\/}.  I have 
taken this deduction seriously, and have considered a conceptual 
scheme in which the notion of absolute observer-independent state 
of a 
system is replaced by the notion of information about a system that 
a 
physical system may possess.

I have considered three postulates that this information must 
satisfy, 
which summarize present experimental evidence about the world.  
The 
first limits the amount of relevant information that a system can 
have; the second summarizes the novelty revealed by the 
experiments 
from which quantum mechanics derives, by asserting that whatever 
the 
information we have about a system we can always get new 
information.  
The third limits the structure of the set of questions; this third 
postulate can probably be sharpened.  Out of these postulates the 
conventional Hilbert space formalism of quantum mechanics and the 
corresponding rules for calculating probabilities (and therefore any 
other equivalent formalism) can be rederived.

A physical system is characterized by the structure on the set 
$W(S)$ 
of questions that can be asked to the system.  This set has the 
structure of the non-Boolean algebra of a family of linear subspaces 
of a complex $k$-dimensional Hilbert space.  The information about 
$S$ 
that any observer $O$ can possess can be represented as a string 
$s$, 
containing an amount of information $N$.

I have investigated the meaning of this information out of which the 
theory is constructed.  I have shown that the fact that a variable in 
a system $O$ has information about a variable in a system $S$ 
means 
that the variables of $S$ and $O$ are correlated, meaning that a 
third 
observer $P$ has information about the coupled $S-O$ system that 
allows her to predict correlated outcomes between questions to $S$ 
and 
questions to $O$.  Thus correlation has no absolute meaning, because 
states have no absolute meaning, and must be interpreted as the 
content of the information that a third system has about the $S-O$ 
couple.

Finally, since we take quantum mechanics as a complete description 
of 
the world at the present level of experimental knowledge 
(hypothesis 
2), we are forced to accept the result that there is no objective, or 
more precisely observer-independent meaning to the ascription of a 
property to a system.  Thus, the properties of the systems are to be 
described by an interrelated net of observations and information 
collected from observations.  Any complex situation can be 
described 
{\it in toto\/} by a further additional observer, and the 
interrelation is consistent.  However, such {\it in toto\/} 
description is deficient in two directions: upward, because an even 
more general observer is needed to describe the global observer 
itself, and --more importantly-- downward, because the {\it in 
toto\/} 
observer knows the content of the information that the single 
component systems possess about each other only probabilistically.

There is no way to ``exit" from the observer-observed global system: 
``Any observation requires an observer'' (The expression is freely 
taken from [Maturana and Varela 1920]).  In other words, I suggest 
that it is a matter of natural science whether or not the 
descriptions 
that different observers give of the same ensemble of events is 
universal or not:

\begin{quote}
{\it Quantum mechanics is the theoretical formalization of the 
experimental discovery that the descriptions that different 
observers 
give of the same events are not universal.}
\end{quote}

The concept that quantum mechanics forces us to give up is the 
concept 
of a description of a system independent from the observer providing 
such a description; that is, the concept of absolute state of a 
system.  The structure of the classical scientific description of the 
world in terms of {\it systems\/} that are in certain states is 
perhaps incorrect, and inappropriate to describe the world beyond 
the 
$\hbar \rightarrow 0$ limit. 

\subsection{Relation with other interpretations}

I conclude with a brief discussion on the relation between the view 
presented here and some popular views of quantum mechanics.  I 
follow 
[Butterfield 1995] to organize current strategies on the quantum 
puzzle.  The first strategy (Dynamics) is to reject the quantum 
postulate that an isolated system evolves according to the linear 
Schr\"odinger equation, and consider additional mechanisms that 
modify 
this evolution --in a sense physically producing the wave function 
collapse.  Examples are the interpretations in which the 
measurement 
process is replaced by some hypothetical process that violates the 
linear Schr\"odinger equation [Ghirardi Rimini and Weber 1986, 
Penrose 
1989].  These interpretation are radically different from the present 
approach, since they violate hypothesis 2.  My effort here is not to 
modify quantum mechanics to make it consistent with my view of 
the 
world, but to modify my view of the world to make it consistent 
with 
quantum mechanics.\footnote{Note added.  I have recently become 
aware 
of an idea to circumvent this problem by exploiting the 
infinite-number-of-degrees-of-freedom nature of the observing 
system.  This could generate an apparently non-linear evolution from 
conventional Schr\"odinger evolution, via a symmetry-breaking 
instability generating effective superselection rules.  See in 
particular [Jona-Lasinio {\it et.  al.} 81, 86] and [Wightman 95].}

The second and third strategies maintain the idea that probabilistic 
expectations of values of any isolated physical system are given by 
the linear Schr\"odinger evolution.  They must then face the problem 
of reconciling (in the example of section II.A) the probabilities 
expressed by the state at time $t_2$ in equation (2), --$q=1$ with 
probability 1/2 and $q=2$ with probability 1/2-- with the assertion 
that the the observer $O$ assigns the value $q=1$ to the variable $q$ 
at the same time $t_2$.  As Butterfield emphasizes, if this value 
assignment coexists with the probability distribution expressed by 
(2), then the eigenstate-eigenvalue link must be in some sense 
weakened, and the possibility of assigning values to variables in 
addition to the eigenstate case (extra values) allowed.  The second 
and the third strategy in Butterfield's classification differ on 
whether these extra values are ``wholly a matter of physics" 
(Physics 
Values), or are ``somehow mental or perspectival" (Perspectival 
Values).  In the first case, the assignment is (in every sense) 
observer-independent.  In the second case, it is (in some sense) 
observer-dependent.

A prime example of the second strategy (Physics Values) is Bohr's, 
or 
Copenhagen, interpretation --at least in one possible reading.  Bohr 
assumes a classical world.  In Bohr's view, this classical world is 
physically distinct from the microsystems described by quantum 
mechanics, and it is precisely the classical nature of the apparatus 
that gives measurement interactions a special status [Bohr 1949; 
for a 
clear discussion of this point, see Landau Lifshitz 1977].

Within the point of view developed in this paper, one can fix once 
and 
for all a privileged system $S_o$ as ``The Observer" (capitalized).  
This system $S_o$ can be formed for instance by all the 
macroscopic 
objects around us.  In this way we recover Bohr's view.  The quantum 
mechanical ``state" of a system $S$ is then the information that the 
privileged system $S_o$ has about $S$.  Bohr's choice is simply the 
assumption of a set of systems (the classical systems) as privileged 
observers.  This is consistent with the view presented 
here.\footnote{A separate problem is why the observing system 
chosen 
--$S_o$, or the macroscopic world-- admits, in turn, a description 
in 
which expectations probabilities evolve classically, namely are 
virtually always concentrated on values 0 and 1, and interference 
terms are invisible.  It is to this question that the physical 
decoherence mechanism [Joos Zeh 1985, Zurek 1981] provides an 
answer.  
Namely, {\it after\/} having an answer on what determines extra-
values 
ascriptions (the observer-observed structure, in the view proposed 
here), the physical decoherence mechanism helps explaining why 
those 
ascriptions are consistent with classical physics in macroscopic 
systems.} By taking Bohr's step, one becomes blind to the net of 
interrelations that are at the foundation of the theory, and puzzled 
about the fact that the theory treats one system, $S_o$, the 
classical 
world, in a way which is different from the other systems.  The 
disturbing aspect of Bohr's view is the inapplicability of quantum 
theory to macrophysics.  This disturbing aspect vanishes, I believe, 
at the light of the discussion in this paper.

Therefore, the considerations in this paper do not suggest any 
modification to the conventional {\it use\/} of quantum mechanics: 
there is nothing incorrect in fixing the preferred observer $S_o$ 
once 
and for all.  If we adopt the point of view suggested here, we 
continue to use quantum mechanics precisely as is it is currently 
used.  On the other hand, this point of view (I hope) brings clarity 
about the physical significance of the strange theoretical procedure 
adopted in Bohr's quantum mechanics: treating a portion of the world 
in a different manner than the rest of it.  This different treatment 
is, I believe, the origin of the unease with quantum mechanics.

The strident aspect of Bohr's quantum mechanics is cleanly 
characterized by von Neumann's introduction of the ``projection 
postulate", according to which systems have two different kinds of 
evolutions: the unitary and deterministic Schr\"odinger evolution, 
and 
the instantaneous, probabilistic measurement collapse [von Neumann 
1932].  According to the point of view described here, the 
Schr\"odinger unitary evolution of the system $S$ breaks down 
simply 
because the system interacts with something which is not taken 
into 
account by the evolution equations.  Unitary evolution requires the 
system to be isolated, which is exactly what ceases to be true 
during 
the measurement, because of the interaction with the observer.  If 
we 
include the observer into the system, then the evolution is still 
unitary, but we are now dealing with the description of a different 
observer.  As suggested by Ashtekar, the point of view presented 
here 
can then be described as a fundamental assumption prohibiting an 
observer to be able to give a full description of ``itself" [Ashtekar 
1993].  In this respect, these ideas are related to earlier 
suggestions that quantum mechanics is a theory that necessarily 
excludes the observer [Peres and Zurek 1982, Roessler 1987, 
Finkelstein 1988, Primas 1990].  A recent result in this regard is a 
general theorem proven by Breuer [Breuer 1994], according to which 
no 
system (quantum nor classical) can perform a complete 
self-measurement.  The relation between the point of view 
presented here and Breuer's result deserves to be explored.

Other views within Butterfield's second strategy (Physical Values) 
are 
Bohm's hidden variables theory, which violates hypothesis 2 
(completeness), and modal interpretations, which deny the collapse 
but 
assume the existence of physical quantities' values.  Of these, I am 
familiar with [van Fraassen 1991], or the idea of actualization of 
potentialities in [Shimony 1969, Fleming 1992].  The assumed values 
must be consistent with the standard theory's predictions, be 
probabilistically determined by a unitary evolving wave function, 
but 
they are not constrained by the eigenstate-eigenvalue link.  One may 
doubt these acrobatics could work [See Bacciagaluppi 1995, 
Bacciagaluppi and Hemmo 1995].  I am very sympathetic with the 
idea 
that the object of quantum mechanics is a set of quantities' values 
and their distribution.  Here, I have assumed value assignment as in 
these interpretations, but with two crucial differences.  First, this 
value assignment is observer-dependent.  Second, it need not be 
consistent with a unitary Schr\"odinger evolution, because the 
evolution is not unitary when the observed system interacts with 
the 
observer.  Namely, there is collapse in each observer-dependent 
evolution of expected probabilities.  These two differences allow 
values to be assigned to physical quantities without any of the 
consistency worries that plague modal interpretations.  The point is 
that the break of the eigenstate-eigenvalue link is bypassed by that 
fact that the eigenvalue refers to one observer, and the state to a 
different observer.  For a fixed observer, the eigenstate-eigenvalue 
link is maintained.  Consistency should only be recovered between 
different observers, but consistency is only quantum mechanical as 
discussed in section III.D. Actuality is observer dependent.  The fact 
that the values of physical quantities are relational and their 
consistency is only probabilistically required circumvents the 
potential difficulties of the modal interpretations.

A class of interpretations of quantum mechanics that Butterfield 
does 
not include in his classification, but which are presently very 
popular among physicists, is the consistent histories (CH) 
interpretations [Griffiths 1984, Omnes 1988, Gell-Mann and Hartle 
1990].  These interpretations reduce the description of a system to 
the prediction of temporal sequences of values of physical variables.  
The key novelties are three: (i) probabilities are assigned to 
sequences of values, as opposed to single values; (ii) only certain 
sequences can be considered; (iii) probability is interpreted as 
probability of the given sequence of values within a chosen family of 
sequences, or framework.  The restriction (ii) incorporates the 
quantum mechanical prohibition of giving value, say, to position and 
momentum at the same time.  More precisely, in combination with 
(iii) 
it excludes all the instances in which observable interference 
effects 
would make probability assignments inconsistent.  In a sense, CH  
represent a sophisticated implementation of the program of 
discovering 
a minimum consistent value attribution scheme.  The price paid for 
consistency is that a single value attribution is meaningless: 
whether 
or not a variable has a value may very well depend on whether we 
are 
asking or not if at a later time another variable has a value.  

There is a key subtlety in the CH scheme that has rarely being 
emphasized\footnote{The only discussion on this point I heard was 
by 
Isham.}: this is the distinction between ``properties that hold with 
probability one'' and ``data''.  The interesting question to ask is: 
What is it that determines which frameworks are allowed?  The 
standard 
answer in CH is that we have certain information on the system, call 
it {\it data\/}.  In particular, we know that certain physical 
quantities of the system have certain values.  The fact that certain 
quantities have certain values determines which framework are 
allowed 
to describe a phenomenon.  Thus: {\it we rely on quantities having 
values (data) for selecting the allowed framework\/}.  Then we 
derive 
probabilistic predictions about value attribution.  These 
probabilistic predictions are framework dependent.  It is often 
stated 
that in CH {\it all\/} value attributions are framework dependent; 
this is the way Nature is.  {\it But if all value attributions are 
framework dependent, what are the data?} In CH there seem to be 
two 
distinct kind of value attributions: a weak value attribution: the 
framework dependent values, and a strong value attribution: the 
data, 
or facts.  In CH the focus is totally on predictions about weak value 
attributions.  But everyday life and scientific practice are about 
(possibly probabilistic) predictions of facts, namely facts that 
could be later used as data.  To put it pictorially (and a bit 
imprecisely): I do not care about a science that tells me that my 
airplane will not crash ``in one framework''; I want a science that 
will tell me that my airplane will just not crash!

Consider the following situation (situation A): a set (D) of data on a 
system is given.  From these data, it follows that there is a 
framework (call it F1) in which the question ``Is $Q$ equal to 1 at 
time t?''  can be asked, and the answer is yes with probability one.  
Then consider the following other situation (situation B): we have a 
set of data on the system, which consist in the set (D) {\it plus the 
data that $Q=1$ at time t\/}.  Is situation A physically the same as 
situation B in the consistent histories approach?  The answer is no.  
In fact, situation A still allows for $Q=2$ (strange but true) in a 
different framework, say F2, while situation B is incompatible with 
$Q=2$, because F2 is not an admitted framework.  {\it Therefore 
there 
should exist a physical way in which we pass from situation A to 
situation B\/}.  Namely there should be way for a probability 1 
prediction (in a framework) to become a data.  The key question that, 
as far as I can see, the consistent histories approaches does not 
address is: how do we concretely pass (in a lab) from situation A to 
situation B? What is it that transforms a probability-one event 
(framework dependent) into something that we can use as data 
(framework independent)?  If the transformation of a 
``probability-one-in-a-framework" situation into {\it data \/} is an 
actual occurrence in Nature, then I believe CH fails to tell me how 
this can happen (when does a framework becomes realized as data).  
If, 
on the other hand, what is data for me may fail to be data for 
somebody other, then one falls precisely into the scheme presented 
in 
this paper.

In the Copenhagen view, it is the interaction with a classical object 
that actualizes properties.  A different solution has been suggested 
in this paper: interaction with any object, but then actualization of 
properties is only relative to that object.  I do not see the solution 
of this problem within the history views.  This is not to say that 
there is anything wrong in the CH approaches.  To the opposite, I 
believe that the CH views are correct and precise.  Still, there is a 
question that they leave open: the physical meaning of the 
framework dependence of the value assignments; more precisely, the 
understanding of how there can be facts, or data, if property 
ascriptions are only framework dependent.  I think that the answer 
is 
simply that there are no (observer independent) data at all: the data 
that I have, and therefore the family of frameworks that I can use is 
different from your data, and therefore the family of framework 
that 
you can use.  The histories interpretations are not inconsistent with 
the analysis developed here.  What I try to add here is attention to 
the process through which the observer-independent, but 
framework-dependent probabilities attached to histories, may be 
related to actual observer-dependent descriptions of the facts of 
the 
world.

Finally, let me come to the third strategy (Perspectival Values), 
whose prime example is the many worlds interpretation [Everett 
1957, 
Wheeler 1957, DeWitt 1970], and its variants.  If the ``branching" of 
the wave function in the many worlds interpretation is considered 
as a 
physical process, it raises the very same sort of difficulties as the 
von Neumann ``collapse" does.  When does it happen?  Which systems 
are 
measuring systems that make the world branch?  These difficulties 
of 
the many-world interpretation have been discussed in the literature 
[See Earman 1986].  Alternatively, we may forget branching as a 
physical process, and keep evolving the wave function under unitary 
evolution.  The problem is then to interpret the observation of the 
``internal" observers.  As discussed in [Butterfield 1985] and [Albert 
1992], this can be done by giving preferred status to special 
observers (apparatus) whose values determine a (perspectival) 
branching.  See Objection 7 in section II.B. A variant is 
to take brains --``Minds"-- as the preferred systems that determine 
this perspectival branching, and thus whose state determines the 
new 
``dimension" of indexicality.  Preferred apparatus, or bringing Minds 
into the game, violates hypothesis 1.

There is a way of having (perspectival) branching keeping all 
systems 
on the same footing: the way followed in this paper, namely to 
assume 
that all values assignments are completely relational, not just 
relational with respect to apparatus or Minds.  Notice, however, that 
from this perspective Everett's wave function is a very misleading 
notion,  not only because it represents the perspective of a 
non-existent observer, but because it even fails to contain any 
relevant 
information about the values observed by each single observer!  
There 
is no description of the universe {\it in-toto\/}, only a 
quantum-interrelated net of partial descriptions.

With respect to Butterfield's classification, the interpretation 
proposed here is thus in the second, as well as in the third, group: 
the extra values assigned are ``somehow perspectival" (but 
definitely 
not mental!), in that they are observer-dependent, but at the same 
time ``wholly a matter of physics", in the sense in which the 
``perspectival" aspect of simultaneity is ``wholly a matter of 
physics" in relativity.  In one word: value assignment in a 
measurement is not inconsistent with unitary evolution of the 
apparatus+system ensemble, because value assignment refers to the 
properties of the system with respect to the apparatus, while the 
unitary evolution refers to properties with respect to an external 
system.

From the point of view discussed here, Bohr's interpretation, 
consistent histories interpretations, as well as the many worlds 
interpretation, are all correct.  The point of view closest to the one 
presented here is perhaps Heisenberg's.  Heisenberg's insistence on 
the fact that the lesson to be taken from the atomic experiments is 
that we should stop thinking of the ``state of the system", has been 
obscured by the subsequent terse definition of the theory in terms of 
states given by Dirac.  Here, I have taken Heisenberg's lesson to  
extreme consequences.\footnote{With a large number of exceptions, 
most physicists hold some version of naive realism, or some version 
of 
naive empiricism.  I am aware of the ``philosophical qualm" that the 
ideas presented here may then generate.  The conventional reply, 
which 
I reiterate, is that Galileo's relational notion of velocity used to 
produce analogous qualms, and that physics seems to have the 
remarkable capacity of challenging even its own conceptual 
premises, 
in the course of its evolution.  Historically, the discovery of 
quantum mechanics has had a strong impact on the philosophical 
credo 
of many physicists, as well as on part of contemporary philosophy.  
It 
is possible that this process is not concluded.  But I certainly do 
not want to venture into philosophical terrains, and I leave this 
aspect of the discussion to competent thinkers.  Just a few 
observations: The relational aspect of knowledge is one of the 
themes 
around which large part of western philosophy has developed.  In 
Kantian terms, only to mention a characteristic voice, any 
phenomenal 
substance which may be object of possible experience is ``entirely 
made up of mere relations" [Kant 1787].  In recent years, the idea 
that the notion of observer- independent description of a system is 
meaningless has become almost a commonplace in disparate areas of 
the 
contemporary culture, from anthropology to certain biology and 
neuro-physiology, from the post-neopositivist tradition to (much 
more 
radically) continental philosophy [Gadamer 1989], all the way to 
theoretical physical education [Bragagnolo Cesari and Facci 1993].  I 
find the fact that quantum mechanics, which has directly 
contributed 
to inspire many of these views, has then remained unconnected to 
these 
conceptual development, quite curious.}

Crane is developing a point of view similar to the one discussed here 
and has attempted an ambitious extension of these ideas to the 
cosmological general-covariant gravitational case [Crane 1995].  It 
was recently brought to my attention that Zurek ends his paper 
[Zurek 
1982] with conclusions that are identical to the ones developed here: 
``Properties of quantum systems have no absolute meaning.  Rather, 
they must be always characterized with respect to other physical 
systems" and ``correlations between the properties of quantum 
systems 
are more basic that the properties themselves" [Zurek 1982].  
Finally, Rob Clifton has brought to my attention an unpublished 
preprint by Kochen [Kochen 1979], with ideas extremely 
similar to the ones presented here.

\vskip.5cm

{\bf Acknowledgments}
\vskip.5cm

The ideas in this work emerged from: i.  Conversations with Abhay 
Ashtekar, Julian Barbour, Alain Connes, J\"urgen Ehlers, Brigitte 
Faulknburg, Gordon Flemming, Jonathan Halliwell, Jim Hartle, Chris 
Isham, Al Janis, Ted Newman, Roger Penrose, Lee Smolin, John 
Wheeler 
and HD Zeh; ii.  A seminar run by Bob Griffiths at Carnegie Mellon 
University (1993); iii.  A seminar run by John Earman at Pittsburgh 
University (1992); iv.  Louis Crane's ideas on quantum cosmology; v.  
The teachings of Paola Cesari on the importance of taking the 
observer 
into account.  It is a pleasure to thank them all.  I also thank 
Gordon Belot, Jeremy Butterfield, Bob Clifton, John Earman, Simon 
Saunders and Euan Squires for discussions and comments on the first
version of this work.

\vskip2cm

{\bf References}
\vskip .5cm

\begin{description}

\item Albert D, 1992, {\it Quantum Mechanics and Experience\/}, 
Cambridge Ma: Harvard University Press.

\item Albert D and Loewer B, 1988, Synthese 77, 195-213; 1989, 
Nous 
23, 169- 186

\item Ashtekar 1993, personal communication

\item Einstein A, Podolsky B, Rosen N 1935, Phys Rev 47

\item Everet H 1957, Rev of Mod Phys 29, 454

\item Bacciagaluppi G 1995, ``A Kochen-Specker Theorem in the 
Modal 
Interpretation of Quantum Theory", International Journal of 
Theoretical Physics, to appear.

\item Bacciagaluppi G and Hemmo M 1995, ``Modal Interpretations of 
Imperfect Measurement", Foundation of physics, to appear.

\item Belifante FJ 1973, {\it A survey of Hidden variable 
theories\/}, 
Pergamon Press

\item Bell J 1987, in {\it Schr\"odinger: centenary of a polymath\/}, 
Cambridge: Cambridge University Press

\item Beltrametti EG, Cassinelli G 1981, {\it The Logic of Quantum 
Mechanics\/}, Addison Wesley

\item Bohm D 1951, {\it Quantum Theory\/}, Englewood Cliffs, NJ

\item Bohr N 1935, Nature 12, 65.

\item Bohr N 1949, discussion with Einstein in {\it Albert Einstein: 
Philosopher-Scientist, \/} Open Court

\item Born M 1926,  Zeitschrift f\"ur Physik 38, 803.

\item Bragagnolo W, Cesari $P$, Facci G 1993, {\it Teoria e metodo 
dell'apprendimento motorio\/}, Bologna: Societa' Stampa Sportiva Ê 

\item Breuer 1994, ``The impossibility of accurate state 
self-measurement", Philosophy of Science, to appear.

\item Butterfield J 1995,{\it Words, Minds and Quanta\/}, in 
``Symposium on Quantum Theory and the Mind" Liverpool

\item Crane L 1995, {\it Clock and Category: Is Quantum Gravity 
Algebraic?  \/} J Math Phys 36

\item D'Espagnat 1971,{\it Conceptual foundations of quantum 
mechanics\/}, Addison-Wesley

\item DeWitt BS 1970, Physics Today 23, 30

\item DeWitt BS, Graham N, 1973 {\it The Many World Interpretation 
of 
Quantum Mechanics\/}, Princeton University Press

\item Dirac PMA 1930, {\it The principles of quantum mechanics\/}, 
Oxford: Clarendon Press

\item Donald M 1990, Proceedings of the Royal Society of London 
A427, 43

\item Earman J 1986, {\it A primer on determinism\/}, Dordrecht: 
Holland Reidel

\item Everett H 1957, Rev of Mod Phys 29, 454

\item Finkelstein D 1969, in {\it Boston Studies in the Philosophy of 
Science\/}, vol 5, eds Cohen RS and Wartofski MW, Dordrecht 1969

\item Finkelstein D 1988, in {\it The universal Turing machine\/}, 
vol 
5, eds R Herken, Oxford University Press

\item Fleming NG 1992, {\it Journal of Speculative Philosophy\/}, 
VI, 
4, 256.
                                                                                             
\item Gadamer HG 1989, {\it Truth and method,\/} New York: 
Crossroad.

\item Gell-Mann M, Hartle J, 1990 in {\it Complexity, Entropy, and 
the 
Physics of Information, SFI Studies in the Sciences of 
Complexity\/}, 
vol III, ed W Zurek.  Addison Wesley

\item Ghirardi GC, Rimini A, Weber T 1986, Phys Rev D34, 470 

\item Greenberger, Horne and Zeilinger 1993, Physics Today

\item Griffiths RB 1984,  J Stat Phys 36, 219; 

\item Griffiths RB 1993, Seminar on foundations of quantum 
mechanics, 
Carnegie Mellon, Pittsburgh, October 1993

\item Griffiths RB 1996 {\it Consistent Histories and Quantum 
Reasoning\/}; quant-ph/9606004.  Phys Rev A, to appear 1996.

\item Halliwel 1994, in {\it Stochastic Evolution of Quantum States 
in 
Open Systems and Measurement Process,\/} L Diosi ed., Budapest.

\item Heisenberg W 1927, Zeit fur Phys 43,172; 1936 {\it Funf 
Wiener 
Vortage\/}, Deuticke, Leipzig and Vienna.

\item Hughes RIG 1989, {\it The structure and interpretation of 
Quantum Mechanics\/}, Harvard University Press, Cambridge Ma

\item Jauch J 1968, {\it Foundations of Quantum Mechanics \/}, 
Adison 
Wesley.

\item Jona-Lasinio G, Martinelli F, Scoppola E 1981, Comm Math 
Phys 
80, 223

\item Jona-Lasinio G, Claverie P 1986, Progr Theor Phys Suppl 86, 
54

\item Joos E and Zee HD, Zeitschrift f\"ur Physik B59, 223

\item Kant E 1787, {\it Critique of Pure Reason\/}, Modern Library 
New 
York 1958

\item Kent A 1995, gr-qc/9512023, to appear in Phys Rev A; 
quant-phys/9511032; gr-qc/9604012; gr-qc/9607073.

\item Kent A and Dowker F 1995, Phys Rev Lett 75, 3038, J Stat 
Phys 
82, 1575

\item Kochen S 1979, {\it The Interpretation of Quantum 
Mechanics\/}, 
unpublished.

\item Landau LD, Lifshitz EM 1977, {\it Quantum Mechanics\/}, 
introduction, Pergamonn Press

\item Lockwood M 1989, {\it Mind brain and the Quantum\/}, Oxford, 
Blackwell

\item Mackey GW 1963 {\it Mathematical Foundations of Quantum 
Mechanics \/} New York, Benjamin

\item Maczinski H 1967, Bulletin de L'Academie Polonaise des 
Sciences 
15, 583

\item Maturana H, Varela F 1980, {\it Autopoiesis and Cognition.  
The 
Realization of the Living\/}, D. Reidel Publishing Company, 
Dordrecht, 
Holland

\item Messiah A 1958, {\it Quantum Mechanics\/},  New York, John 
Wiley

\item Newman ET 1993, Talk at the inaugural ceremony of the 
Center for 
Gravitational Physics, Penn State University, August 1993

\item Omnes R 1988, J Stat Phys 57, 357

\item Penrose R 1989, {\it The emperor's new mind\/}, Oxford 
University Press

\item Peres A, Zurek WH 1982, American Journal of Physics 50, 807

\item Piron C 1972, Foundations of Physics 2, 287

\item Primas H 1990, in {\it Sixty-two years of uncertainty\/}, eds 
AI 
Miller, New York Plenum

\item Roessler OE 1987, in {\it Real brains - Artificial minds\/}, 
eds 
JL Casti A Karlqvist, New York, North Holland

\item Rovelli C 1995,  ``Half way through the woods'',
in {\it The Cosmos of Science}, J Earman and JD Norton editors,
(University of Pittsburgh Press / Universitaets Verlag Konstanz,
1997), in print

\item Shannon CE 1949, {\it The mathematical theory of 
communication\/} , University of Illinois Press

\item Shimony A 1969, in {\it Quantum Concepts and spacetime,\/} 
eds R 
Penrose C Isham, Oxford: Clarendon Press.

\item Schr\"odinger E 1935, Naturwissenshaften 22, 807

\item Van Fraassen B 1991, {\it Quantum Mechanics: an Empiricist 
View\/}, Oxford University Press

\item Von Neumann J 1932, {\it Mathematische Grundlagen der 
Quantenmechanik\/}, Springer, Berlin 

\item Wheeler JA 1957, Rev of Mod Phys 29, 463

\item Wheeler JA 1988, {\it IBM Journal of Research and 
Development,\/} Vol 32, 1

\item Wheeler JA 1989, {\it Proceedings of the 3rd International 
Symposium on the Foundations of Quantum Mechanics\/}, Tokyo

\item Wheeler JA 1992, {\it It from Bit and Quantum Gravity\/} , 
Princeton University Report

\item Wheeler J, Zurek W 1983, {\it Quantum Theory and 
Measurement 
\/} Princeton University Press

\item Wightman AS 1995, in {\it Mesoscopic Physics and 
Fundamental 
Problems in Quantum Mechanics\/}, edited by E Di Castro, F Guerra, G 
Jona-Lasinio.
 
\item Wigner EP 1961, in {\it The Scientist Speculates\/} , ed Good, 
New York: Basic Books.

\item Zurek WH 1981, Phys Rev D24 1516

\item Zurek WH 1982, Phys Rev D26 1862

\end{description}
\end{document}